\documentclass[12pt,preprint]{aastex}

\usepackage{color}
\usepackage{graphicx}
\usepackage{float}

\newcommand{\degree}{\ensuremath{^\circ}}
\newcommand{\hrrl}[1]{H{#1}$\alpha$}
\def\ga{\mathrel{\raise0.35ex\hbox{$\scriptstyle >$}\kern-0.6em
\lower0.40ex\hbox{{$\scriptstyle \sim$}}}}
\def\la{\mathrel{\raise0.35ex\hbox{$\scriptstyle <$}\kern-0.6em
\lower0.40ex\hbox{{$\scriptstyle \sim$}}}}
\def\co{CO {J}=1--0 }

\def\ee #1 {\times 10^{#1}}          
\def\ut #1 #2 { \, \mathrm{#1}^{#2}} 
\def\u #1 { \, \mathrm{#1}}          
\def\msol{\hbox{$\hbox{M}_\odot$}}

\begin{document}

\title{The Science Cases for Building \\
a Band 1 Receiver Suite for ALMA}

\author{
J.~Di~Francesco\altaffilmark{1,2},
D.~Johnstone\altaffilmark{1,2},
B.~Matthews\altaffilmark{1,2},
N.~Bartel\altaffilmark{3},
L.~Bronfman\altaffilmark{4},
S.~Casassus\altaffilmark{4},
S.~Chitsazzadeh\altaffilmark{2,5},
H.~Chou\altaffilmark{6},
M.~Cunningham\altaffilmark{7},
G.~Duch\^{e}ne\altaffilmark{8,9},
J.~Geisbuesch\altaffilmark{10},
A.~Hales\altaffilmark{11},
P.T.P.~Ho\altaffilmark{6}
M.~Houde\altaffilmark{5},
D.~Iono\altaffilmark{12},
F.~Kemper\altaffilmark{6},
A.~Kepley\altaffilmark{11},
P.M.~Koch\altaffilmark{6},
K.~Kohno\altaffilmark{13},
R.~Kothes\altaffilmark{10},
S-P.~Lai\altaffilmark{14},
K.Y.~Lin\altaffilmark{6},
S.-Y.~Liu\altaffilmark{6},
B.~Mason\altaffilmark{11},
T.J.~Maccarone\altaffilmark{15},
N.~Mizuno\altaffilmark{12},
O.~Morata\altaffilmark{6},
G.~Schieven\altaffilmark{1},
A.M.M.~Scaife\altaffilmark{16},
D.~Scott\altaffilmark{17},
H.~Shang\altaffilmark{6},
M.~Shimojo\altaffilmark{12},
Y.-N.~Su\altaffilmark{6},
S.~Takakuwa\altaffilmark{6},
J.~Wagg\altaffilmark{18,19},
A.~Wootten\altaffilmark{11},
F.~Yusef-Zadeh\altaffilmark{20}
}

\altaffiltext{1}{National Research Council Canada, 5071 West Saanich Rd, Victoria, BC, V9E 2E7, Canada}
\altaffiltext{2}{Dept.\ of Physics \& Astronomy, University of Victoria, Victoria, BC, V8P 1A1, Canada}
\altaffiltext{3}{Dept.\ of Physics and Astronomy, York University, Toronto, M3J 1P3, ON, Canada}
\altaffiltext{4}{Dept.\ de Astronom\'{\i}a, Universidad de Chile, Casilla 36-D, Santiago, Chile}
\altaffiltext{5}{Dept.\ of Physics and Astronomy, The University of Western Ontario, London, ON, N6A 3K7, Canada}
\altaffiltext{6}{Academia Sinica, Institute of Astronomy and Astrophysics, P.O.\ Box 23-141, Taipei 10617, Taiwan}
\altaffiltext{7}{School of Physics, University of New South Wales, Sydney, NSW 20152, Australia}
\altaffiltext{8}{Astronomy Dept., University of California, Berkeley, CA 94720-3411, USA}
\altaffiltext{9}{Universit\'e Joseph Fourier - Grenoble 1/CNRS, LAOG UMR 5571, BP 53, 38041 Grenoble, France}
\altaffiltext{10}{National Research Council Canada, P.O.\ Box 248, Penticton, BC, V2A 6J9, Canada}
\altaffiltext{11}{National Radio Astronomy Observatory, 520 Edgemont Road, Charlottesville, Virginia 22903, USA}
\altaffiltext{12}{National Astronomical Observatory of Japan, 2-21-1 Osawa, Mitaka, Tokyo, 181-8588, Japan}
\altaffiltext{13}{Institute of Astronomy, The University of Tokyo, 2-21-1 Osawa, Mitaka,Tokyo 181-0015, Japan}
\altaffiltext{14}{Institute of Astronomy and Dept.\ of Physics, National Tsing Hua University, Taiwan}
\altaffiltext{15}{Department of Physics, Texas Tech University, Lubbock, TX, 79409-1051, USA}
\altaffiltext{16}{School of Physics and Astronomy, University of Southampton, Southampton, Hampshire, S017 1BJ, UK}
\altaffiltext{17}{Dept.\ of Physics and Astronomy, University of British Columbia, Vancouver, BC, V6T 1Z1, Canada}
\altaffiltext{18}{European Southern Observatory, Alonso de Cordova 3107, Vitacura, Casilla 19001, Santiago 19, Chile}
\altaffiltext{19}{Astrophysics Group, Cavendish Laboratory, JJ Thomson Avenue, Cambridge, CB30HE, UK}
\altaffiltext{20}{Dept.\ of Physics and Astronomy and Center for Interdisciplinary Research in Astronomy, Northwestern 
University, Evanston, IL 60208, USA}

~
\section{Executive Summary}
\label{sec:exec}

The ALMA Band 1 project aims to provide a low-cost solution to one of the original design goals of the Atacama
Large Millimeter/submillimeter Array (ALMA), access to frequencies of $\sim$40 GHz at high resolution and
sensitivity from the southern hemisphere.  In this document, we present a set of compelling science cases for
construction of the ALMA Band 1 receiver suite.  For these, we assume in tandem the updated nominal Band 1
frequency range of 35-50 GHz and its likely extension up to 52 GHz that together optimize the Band 1 science
return.  

A comprehensive comparison of ALMA and the Jansky VLA (JVLA) over 40-50 GHz finds ALMA having similar
sensitivity at lower frequencies but the edge in sensitivity (e.g., up to a factor of $\sim$2) at higher frequencies.
In addition, ALMA's larger primary beams allow this sensitivity to be obtained over wider fields.  Furthermore,
ALMA Band 1 images will have significantly greater fidelity than those from the JVLA since ALMA has a larger
number of instantaneous baselines.  ALMA's smaller dishes (and the ACA, if needed) in principle can allow the
recovery of more extended emission.  Finally, ALMA Band 1 will likely include frequencies of 50-52 GHz that 
the JVLA simply cannot observe.  

The scope of the science cases ranges from nearby stars to the re-ionization edge of the Universe.  Two cases
provide additional leverage on the present ALMA Level One Science Goals and are seen as particularly powerful
motivations for building the Band 1 receiver suite: (1) detailing the evolution of grains in protoplanetary disks,
as a complement to the gas kinematics, requires continuum observations out to 35 GHz ($\sim$9 mm); and (2)
detecting CO 3--2 line emission from galaxies like the Milky Way during the epoch of re-ionization, i.e., 6 $< z <$
10, also requires Band 1 receiver coverage.   Indeed, Band 1 will increase the volume of the observable Universe
in CO lines by a factor of 8.  The range of Band 1 science is very broad, however, and also includes studies of
galaxy clusters (i.e., via the Sunyaev-Zel'dovich Effect), very small dust grains in the ISM, the Galactic Center,
solar studies, pulsar wind nebulae, radio supernovae, X-ray binaries, dense cloud cores, complex carbon-chain
molecules, ionized gas (e.g., in HII regions), masers, magnetic fields in the dense ISM, jets and outflows from
young stars, the co-evolution of star formation with active galactic nuclei, and the molecular mass in moderate
redshift galaxies.

\newpage

\section{Introduction}
\label{sec:intro}

The Atacama Large Millimeter/submillimeter Array (ALMA) will be a single research instrument composed of at least
fifty 12-m antennas in its 12-m Array and twelve 7-m high-precision antennas plus four 12-m antennas in its compact
array (the Atacama Compact Array; ACA), located at a very high altitude of 5000 m on the Chajnantor plateau of the
Chilean Andes.  The weather conditions at the ALMA site will allow transformational research into the
physics of the cold Universe across a wide range of wavelengths, from radio to submillimeter.  Thus, ALMA will be
capable of probing the first stars and galaxies and directly imaging the disks in which planets are formed.  ALMA will
be the pre-eminent astronomical imaging and spectroscopic instrument at millimetre/submillimetre wavelengths for
decades to come.  It will provide scientists with capabilities and wavelength coverage that complement those of other
key research facilities of its era, such as the James Webb Space Telescope (JWST), 30-m class Giant Segmented
Mirror Telescopes (GSMTs), and the Square Kilometer Array (SKA).  

ALMA will revolutionize many areas of astronomy and an amazing breadth of science has already been proposed (see,
for example, the ALMA Design Reference Science Plan). The technical requirements of the ALMA Project are, however, 
driven by three specific Level One Science Goals:

\noindent{\bf (1)} The ability to detect spectral line emission from CO or CII in a normal galaxy like the Milky Way at a 
redshift of $z = 3$, in less than 24 hours of observation.\\
\noindent{\bf (2)} The ability to image the gas kinematics in a solar-mass protostellar/protoplanetary disk at a distance of 
150 pc (roughly the distance of the star-forming clouds in Ophiuchus or Corona Australis), enabling one to study the 
physical, chemical, and magnetic field structure of the disk and to detect the tidal gaps created by planets undergoing 
formation.\\
\noindent{\bf (3)} The ability to provide precise images at an angular resolution of 0.1$^{\prime\prime}$. Here the term 
``precise image" means an accurate representation of the sky brightness at all points where the brightness is greater than 
0.1\% of the peak image brightness. This requirement applies to all sources visible to ALMA that transit at an elevation 
greater than 20\degree.

ALMA was originally envisioned to provide access to all frequencies between 31 GHz and 950 GHz  accessible from the 
ground.  During a re-baselining exercise undertaken in 2001, the entire project was scrutinized to find necessary cost 
savings.  The two lowest receiver frequencies, Bands 1 and 2, covering 31--45 GHz and 67--90 GHz respectively, were 
among those items delayed beyond the start of science operations. Nevertheless, Band 1 was re-affirmed as a high priority 
future item for ALMA.

In May 2001, John Richer and Geoff Blake prepared the document {\it Science with Band 1 (31--45 GHz) on ALMA}
as part of the re-baselining exercise.  Key arguments for Band 1 receivers included their abilities to: (1) enable exciting 
science opportunities, bringing in a wider community of users; (2) be a significantly faster imaging and survey instrument 
than the upgraded VLA (now known as the Jansky VLA or JVLA), especially due to the larger primary beam; (3) provide
access to the southern sky at these wavelengths; (4) allow excellent science possible even in ``poor" weather; (5) be a 
relatively cheap and reliable receiver to build and maintain; and (6) be a very useful engineering/debugging tool for the 
entire array given the lower contribution of the atmosphere at many of its frequencies relative to other bands.

The Richer/Blake document was followed by an ASAC Committee Report in October 2001, after the addition of Japan
into the ALMA project re-opened the question of observing frequency priorities for those Bands which had been put on 
hold during re-baselining. The unanimous recommendation of the ASAC was to put Band 10 as top priority, followed by
a very high priority Band 1. At that time, the key science cases for Band 1 receivers were seen to be (1) high-resolution 
Sunyaev-Zel'dovich effect (SZE) imaging of cluster gas at all redshifts; and (2) mapping the cold ISM in Galaxies at 
intermediate and high redshift.

The scientific landscape has changed significantly since 2001 and thus it is time to re-examine the main science 
drivers for ALMA Band 1 receivers, even reconsidering the nominal frequency range of Band 1 itself to optimize the
science return.  In addition, the ALMA Development process has begun, and now is the time to put forth the best case
for longer wavelength observing with ALMA.  In October 2008, two dozen astronomers from around the globe met in 
Victoria, Canada to discuss Band 1 science.  This paper summarizes the outstanding cases made possible with Band
1 that were highlighted at that meeting and since.  In Section \ref{sec:freq}, we describe the new nominal Band 1
frequency range of 35-50 GHz, and its likely extension to 52 GHz.
In Section \ref{sec:one}, we present two science cases that reaffirm and enhance the already established ALMA
Project Level One Science Goals. Section \ref{sec:suitability} discusses both weather
considerations at the ALMA site and compares the observing capabilities of ALMA and the JVLA over Band 1 frequencies.
  In Section \ref{sec:range}, we provide a selection of continuum and line science
cases that reinforce the breadth and versatility of the Band 1 receiver suite.  Finally, Section \ref{sec:conc} briefly
summarizes the report.
\newpage

\section{The Band 1 Frequency Range}
\label{sec:freq}

Band 1 was originally defined as 31.3--45 GHz, with the lower end set to the lower edge of a frequency range 
assigned to radio astronomy and the upper set to include SiO $J$=1--0 emission at 43 GHz.  Receiver technology
advances, however, have made it possible to widen and shift the Band 1 range and optimize the science
return of Band 1.  For example, a wider range and shift to higher frequencies will allow molecular emission from
galaxies at a wider range of (slightly lower) redshifts to be explored.  Also, it will allow molecular emission from
several new species in our Galaxy to be probed.  (Of course, this shift does in turn remove the ability to detect
molecular emission from some higher redshift galaxies and some other Galactic transitions.)  Furthermore, a
shift to higher frequencies for Band 1 will improve (slightly) the angular resolution of continuum observations
and better exploit the advantages of the dry ALMA site.  

A review of the nominal frequency range by the Band 1 Science Team (i.e., several authors of this document) in June 2012
resulted in a proposed new Band 1 frequency range definition, nominally 35--50 GHz with a likely extension up to 52 GHz.
The shift up to 50 GHz will allow the important line CS $J$=1--0 at 48.99 GHz to be observable with ALMA. In addition, the
nominal range of 35-50 GHz alone is itself $\sim$10\% wider than before.  As it will provide the highest sensitivities, the
nominal range will be preferred for high-redshift science.  The extension to 50-52 GHz, which the JVLA cannot observe, may
be somewhat adversely affected by atmospheric O$_{2}$, resulting in lower relative sensitivity.  Since numerous transitions
of other interesting molecules have rest frequencies at 50-52 GHz, however, this extension will allow such emission to be
probed toward sources in our Galaxy.  This document has been updated in September 2012 to reflect the new nominal frequency
range and the extension.  See Section~\ref{sec:suitability} for a comparison of the sensitivities and imaging
characteristics of ALMA and the JVLA over Band 1 frequencies.

\newpage
\section{Level One Science Cases for Band 1}
\label{sec:one}

In this section, we present two science cases that reaffirm and enhance the already established ALMA Project Level One
Science Goals: Evolution of Grains in Disks Around Stars (\S\ref{sec:disk}) and The First Generation of Galaxies
(\S\ref{sec:red}).  Further science cases are presented in \S6.

\subsection{Evolution of Grains in Disks Around Stars}
\label{sec:disk}
\subsubsection{Protoplanetary Disks}

Planet formation takes place in disks of dust and gas surrounding
young stars. It is within these gas-rich protoplanetary disks that
dust grains must agglomerate from the sub-micron sizes associated with
the interstellar medium to larger pebbles, rocks and planetesimals, if
planets are ultimately to be formed.
The timescale of this agglomeration process is
thought to be a few tens of Myr for terrestrial planets, while the
process leading to the formation of giant planet cores remains
uncertain. Core accretion models require at least a few Myr to form
Jovian planets (Pollack et al.\ 1996), while dynamical instability
models could form giant planets on orbital timescales ($t \ll 1$ Myr;
Boss 2005).

\begin{figure}
\begin{center}
\includegraphics[scale=0.55]{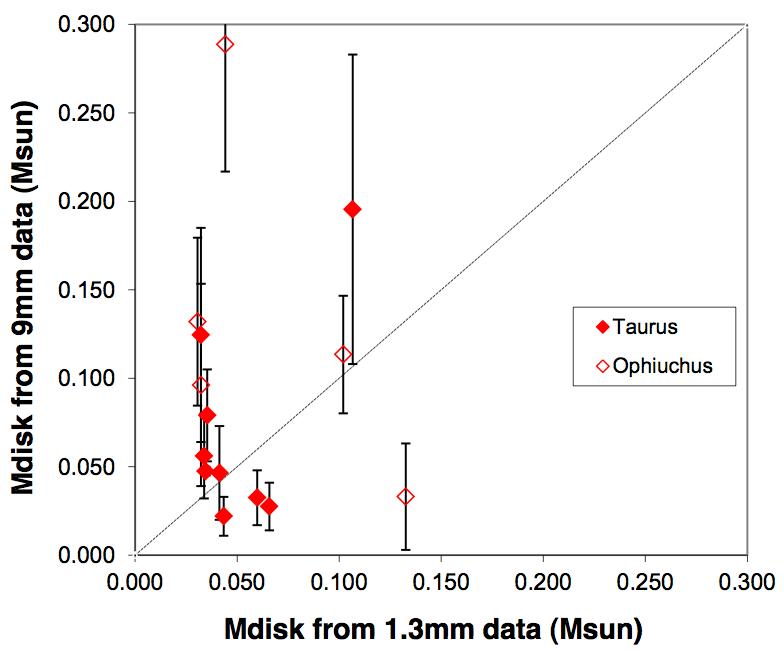}
\caption{Disk masses measured from 9 mm continuum emission compared to those measured from 1.3 mm continuum emission in the regions of Taurus and Ophiuchus.  Many disks show higher mass measurements at the longer wavelength, indicating the presence of larger grains than those detected at 1.3 mm measurements. (Greaves et al., in prep.)}
\label{fig:diskmasses}
\end{center}
\end{figure}

Gravitational instability models require high disk masses to
form planets. So far, most accurate disk mass estimates come from
submillimeter and millimeter
observations, where the dust is optically thin. Andrews
\& Williams (2007a, 2007b) show that submillimeter
observations of dozens of protoplanetary disks reveal that only
one system could be gravitationally unstable, conflicting with the
high frequency of Jovian planets seen around low mass stars. Have
these relatively young (1--6 Myr) systems already formed planets, or is
most of the dust mass locked into larger grains and therefore not
accounted for in  submillimeter and millimeter observations?  If grain growth to
centimeter sizes has occurred, most of a disk's dust mass would reside in
the large particle population, which would emit at longer millimeter and 
centimeter wavelengths. Figure \ref{fig:diskmasses} from Greaves et al.\
(in prep.)\
compares disk masses for
objects in Taurus and Ophiuchus derived from 9 mm and 1.3 mm  dust
fluxes. The longer wavelength masses are found to be generally higher
than the shorter wavelength values, indicating that a significant
fraction of the disks' total dust masses are indeed locked up in
larger grains.

Identifying {\it where} and {\it when} dust coagulation occurs is
critical to constrain current models of planetary formation. Growth
from sub-micron to micron-sized particles can be traced with infrared
spectroscopy and imaging polarimetry. The next step, growth beyond 
micron sizes, is readily studied by determining the slope of the
spectral energy distribution (SED) of the dust thermal emission at
 submillimeter and millimeter wavelengths. The dust mass opacity index at wavelengths 
longer than 0.1 millimeter is
approximately a power-law whose normalization depends on the dust
properties, such as composition, size distribution, and geometry
(Draine 2006). The index of the power law is commonly given by
$\beta$.  The presence of large grains is detectable through a decrease
in $\beta$,
which can be derived directly from the slope of the Rayleigh-Jeans
tail of the SED, $\alpha$, where $\beta = \alpha -2$ when the emission is optically thin. 
Numerous studies have revealed that the $\beta$ values of disks are
substantially lower than the typical ISM value of $\sim 2$ (e.g., Testi et
al.\ 2003; Weintraub et al.\ 1989; Adams et al.\ 1990; Beckwith et al.\
1990; Beckwith \& Sargent 1991; Mannings \& Emerson 1994).

The key stumbling block to the interpretation of $\beta$ occurs when
the disk is not resolved spatially.  The amount of flux detected at a
given wavelength is a function of both $\beta$ and the size of the
disk (Testi et al.\ 2001). Resolving the ambiguity therefore is truly a
matter of resolution, and sufficient resolution is only offered at
these wavelengths by interferometers.

Among the three high level science goals of ALMA is the ability to
detect and image gas kinematics in protoplanetary disks undergoing
planetary formation at 150~pc.  At ALMA's observing wavelengths, its
capability for imaging the continuum dust emission in these disks is
also second-to-none.  At present, however, the longest wavelength 
that ALMA can reach is 3.6 mm.  Given that dust particles emit very
inefficiently at wavelengths longer than their sizes, the present
ALMA design will not be sensitive to particles larger than $\sim 3$ mm.
{\it This situation negates ALMA's potential ability to follow the dust grain
growth from mm-sized to cm-sized pebbles in protoplanetary disks.}

\begin{figure}
\begin{center}
\includegraphics[scale=0.65]{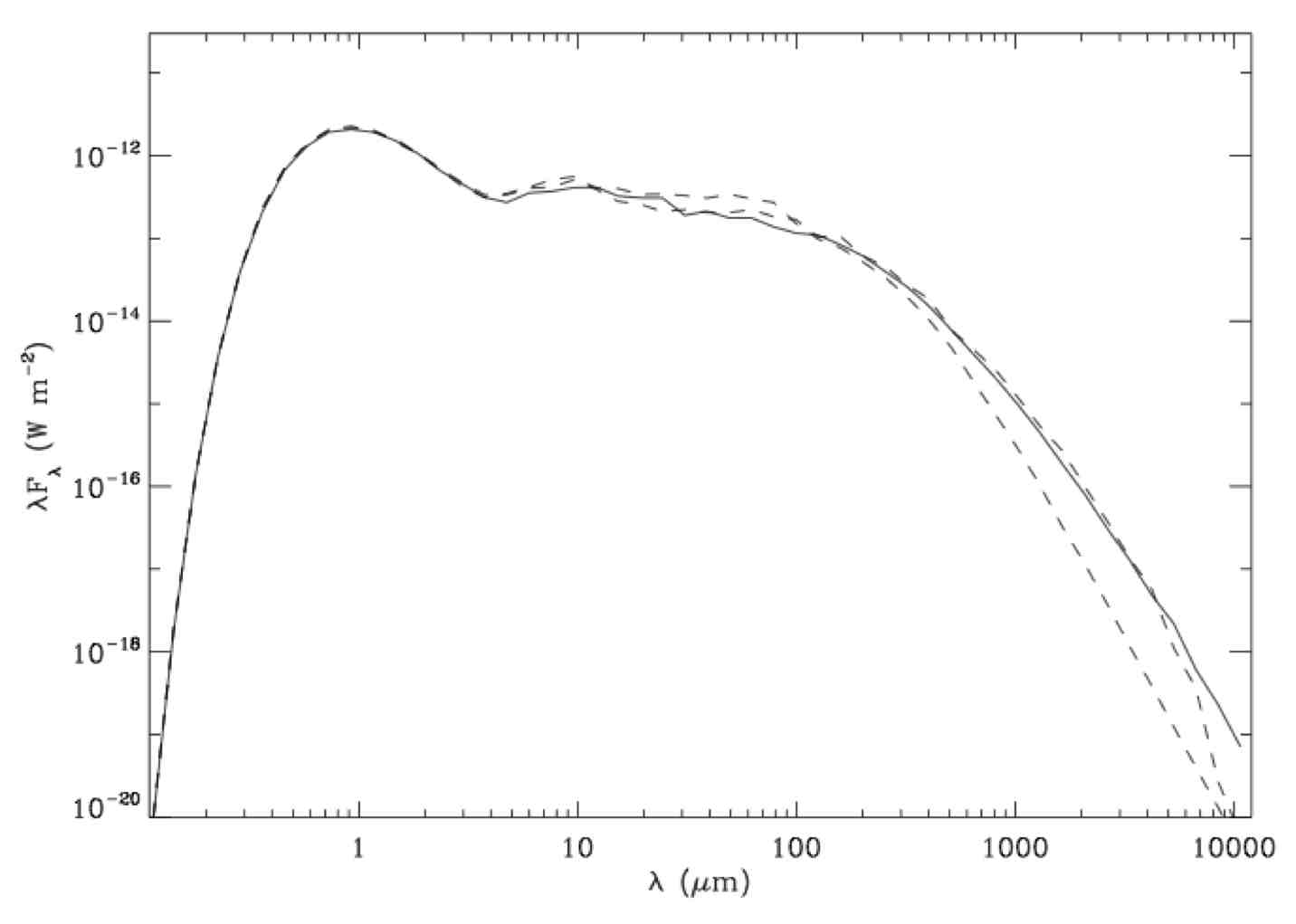}
\caption{Spectral energy distribution plot showing the differences
between three disk models having different maximum grain sizes.  The
solid curve is the model with a$_{\rm max}=1 $ cm, which keeps declining with
roughly constant slope all the way to 1 cm. The two dashed curves are
for a$_{\rm max}=10\ \mu$m and 1 mm. The top one, which breaks around 5 mm is the
model with a$_{\rm max}=1 $ mm. It's interesting to note how the fluxes are very
much the same for a$_{\rm max}= 1 $ mm or 1 cm, except precisely towards ALMA's
Band 1.  There is at least an order of magnitude difference in power
at 1 cm between the max$_{\rm size}=1 $ mm versus the max$_{\rm size}=1 $ cm disks. These
models indicate that observations at the ALMA Band 1 regime are crucial
for determining whether grain-growth to cm-sizes is indeed occurring.}
\label{fig:disk1}
\end{center}
\end{figure}

Figure~\ref{fig:disk1} shows the SEDs for three different circumstellar disk
models, computed using the full dust radiative transfer MCFOST code
(Pinte et al.\ 2006; Pinte et al.\ 2009). The model parameters are
representative of protoplanetary disks (although there is substantial
object-to-object variation). The circumstellar disk is passively
heated by a 4000 K, 2 L$_{\odot}$ central star and the system is located
160 pc away. The dust component of the disk is assumed to be fully
mixed with the gas and the latter is assumed to be in vertical
hydrostatic equilibrium. The disk extends radially from 1 AU to
100 AU. The total dust mass in the model is $10^{-3}$ M$_\odot$ (the gas
component is irrelevant for continuum emission calculations, so its
mass is not set in the model, though a typical 100:1 gas:dust ratio is
generally assumed). The dust population is described by a single power-law
size distribution $N(a)\propto a^{-3.5}$ with a minimum grain size of 0.03
$\mu$m and extending to 10 $\mu$m, 1 mm or 1 cm depending on the
model. The dust composition is taken to be the ``astronomical
silicates" model from Draine (2003).

Figure~\ref{fig:disk1} reveals that observations in the ALMA Band 1 spectral
region are crucial for determining whether grain-growth to cm-sizes is
indeed occurring. The 1 cm flux density of the max$_{size}=$1\,cm disk model is
$\sim50\,\mu$Jy, comparable to the 1 $\sigma$ sensitivities provided by
ALMA's Band 1 with 1 minute integration. Besides ALMA, there are no
existing or planned southern astronomical facilities capable of
observing to such depths at these frequencies. Therefore, {\it if ALMA
Band 1 receivers are not built there will be no way of putting ALMA
observations of protoplanetary disks in the context of coagulation of
dust grains to centimeter sizes.}  Though such information could be 
acquired in part with the JVLA (for sufficiently northern sources), ALMA
Band 1 receivers would yield superior data for comparison with those
of other Bands, given greater similarities in spatial frequency coverage.
(Spatial frequency coverage depends on the latitude of the observatory
and the declination of the source.)

\begin{figure}[h]
\begin{center}
\includegraphics[scale=0.65]{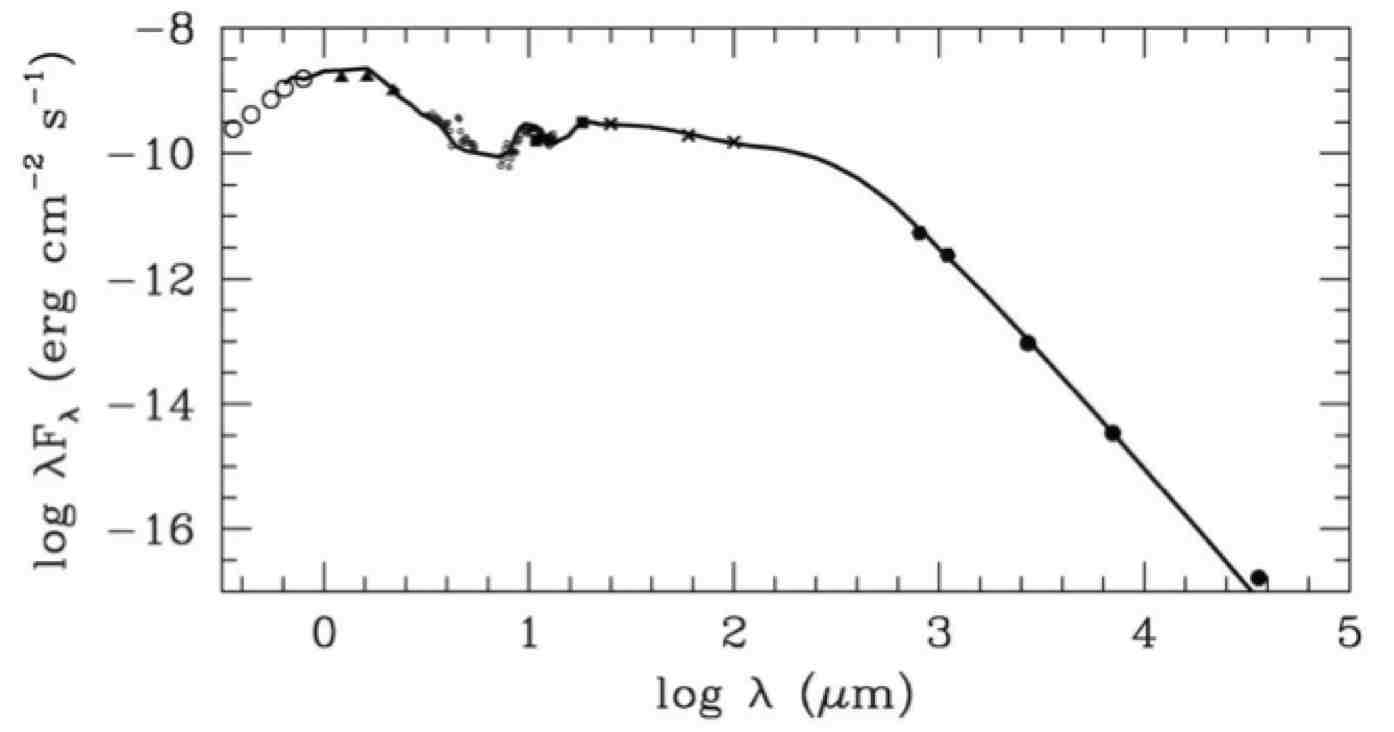}
\caption{Spectral energy distribution of TW Hya, showing
the fit to the SED for an irradiated accretion disk model with a maximum
particle size of 1 cm (Calvet et al.\ 2002).}
\label{fig:disk2}
\end{center}
\end{figure}

By complementing observations in other ALMA Bands, Band 1 will provide a
crucial longer wavelength lever to minimize the uncertainty in
$\alpha$.  Evidence for small pebbles has
been detected in several disks (Rodmann et al.\ 2006). The prime
example is TW Hya, a protoplanetary disk 50 pc from the Sun (Wilner et
al.\ 2000).  Its SED is well matched by an irradiated accretion disk
model fit from 10s of AU to an outer radius of 200 AU and requires the
presence of particle sizes up to 1 cm in the disk (see Figure~\ref{fig:disk2}). 
The measured $\beta$ is $0.7 \pm 0.1$ (Calvet et al.\ 2002).  To date,
no trend in $\beta$ has been detected with stellar luminosity, mass or
age (Ricci et al. 2010). Lower $\alpha$ values are associated with less
60 $\mu$m excess,
however, suggesting that settling or agglomeration processes could be
removing the smallest grains, decreasing the shorter wavelength
emission (Acke et al.\ 2004).  (See \S6.1.1 for further discussion of 
probes of small grains with the ALMA Band 1 receivers.)

\begin{figure}[h]
\begin{center}
\includegraphics[scale=0.65]{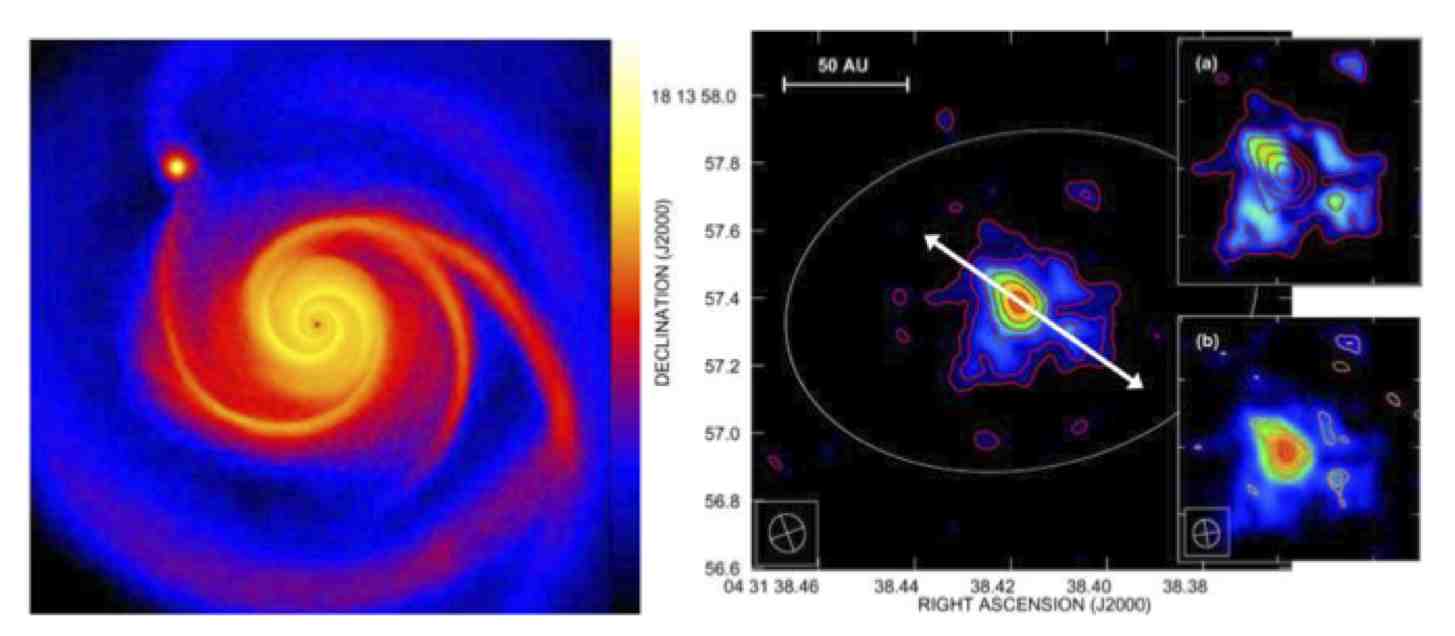}
\caption{(Left) Image from an SPH simulation showing the surface
density structure of a 0.3~M$_\odot$ disk around a 0.5~M$_\odot$ star. A
single dense clump has formed in the disk, at a radius of 75 AU and
with a mass of $\sim 8$ M$_{\rm Jup}$. {(Right)} VLA 1.3 cm images toward HL
Tau. The main image shows natural weighting with a beam of
0.11$^{\prime\prime}$ FWHM. The arrow indicates the jet axis. Upper inset:
compact central peak subtracted. Lower inset: uniform weighting, with a
beam of 0.08$^{\prime\prime}$ FWHM.  The compact object lies to the upper
right hand side. This condensation was also detected at 1.4 mm with the
BIMA array (Welch et al.\ 2004).}
\label{fig:disk3}
\end{center}
\end{figure}

At the resolution provided by its longest baselines at $\sim$40 GHz
($\sim$0.14$^{\prime\prime}$), ALMA will easily resolve protoplanetary disks
at the distance of the closest star-forming regions (50--150 pc).  These
resolved images will provide the most accurate determination of the
disk's dust mass.  The dust distribution at centimeter wavelengths can then be
compared to millimeter and submillimeter images, revealing where in the disk 
dust coagulation is occurring.  For example, previous investigations of the
radial dependency of dust properties in disks by Guilloteau et al.\ (2009) and
Isella et al.\ (2010) were conducted at 1 mm and 3 mm, and as such they were
sensitive to only millimeter-sized grains.  Note, however, that Melis et al.\ 
(2011) used the Jansky VLA to map the 7 mm emission from the protoplanetary
disk around the young source L1527 IRS at $\sim$1.5$^{\prime\prime}$ and
tentatively detected a {\it dearth}\/ of ``pebble-sized" grains. ALMA Band 1
receivers will help clarify this situation. 
As described above, Band 1 data will be sensitive to larger grains.  Moreover,
through detection of concentrations of such large grains, protoplanets in formation
can be identified. These condensations are predicted by simulations of gravitational
instability models (see Figure~\ref{fig:disk3}a; Greaves et al.\ 2008) and have been
detected in the nearby star HL Tau (Figure~\ref{fig:disk3}b; Greaves et al.\ 2008).

Detecting dust emission at centimeter wavelengths also requires high sensitivity,
because its brightness is several orders of magnitude lower than in the
submillimeter.  In addition, at wavelengths longer than 7 mm (i.e., $\nu$ $<$ 
45 GHz), the contribution
from other radiative processes, such as ionized winds, can contribute
significantly to the total flux and complicate the interpretation of
detected emission. 
Rodmann et al.\ (2006) found that the contribution of free-free
emission to the total flux is typically 25\% at a wavelength of 7
mm.  Observations of continuum emission at the 35-50 GHz
(6--9 mm) spectral
range enabled by Band 1 would increase substantially the sampling
rate in the region where emission is detected from both the free-free
and thermal dust emission components.  
The synergy with the JVLA will provide a longer wavelength lever for sources
observed in common, providing an estimate of the free-free contribution to the
Band 1 flux.  Such data would not be essential, however, given wide frequency
coverage within Band 1 alone.  For example, multiple continuum observations
could be used to quantify accurately the relative amounts of free-free and dust
emission through changes in spectral slope, and thereby determine precisely
the contribution from large dust grains (i.e., protoplanetary material).

In summary, ALMA Band 1 receivers would provide the sensitivity to long
wavelength emission needed to probe dust coagulation and growth in 
protoplanetary disks observed at higher-frequency bands.  Of course, 
ALMA Band 1 will allow such investigations of sources too far south to
observe with the JVLA.  (For the highest resolutions, the improved phase
stability available at ALMA will also be very important.)  Furthermore, as
comparisons with higher frequencies are better when there is similar
spatial frequency coverage, however, sources are best observed at
different wavelengths from the same latitude, favouring ALMA data
over JVLA data even for northern sources.

\subsubsection{Debris Disks}

Around main sequence stars, pebble-sized bodies are produced differently 
than in disks around pre-main-sequence stars.  Here, destructive collisional
cascades from even larger planetesimals through to centimeter, millimeter,
and then micron-sized particles provides ongoing replenishment of the debris
population (Wyatt 2009; Dullemond \& Dominik 2005).  The methods for
detecting large (i.e., centimeter-sized) grains is the same as in
protoplanetary disks, despite their origin in destructive rather than
agglomerative processes. In each case, the longer the wavelength at 
which continuum emission is detected, the larger the grains that must be 
present in the system.

Using ALMA Band 7, Boley et al.\ (2012) detected the debris disk of Fomalhaut,
and noted its sharp inner and outer boundary.  Band 1 images, however,
could show higher contrast features in debris disks compared to other ALMA
Bands, due to the longer resonant lifetimes of
the larger particles that dominate the emission.  This sensitivity in turn will
help detect any edges and gaps in the disks. 
Dramatic changes in the morphology of debris disks as a function of wavelength 
have already been observed (e.g., Maness et al.\ 2008), but not yet at 
the long wavelengths Band 1 will probe. When observed,
such structures are often considered signposts to the existence of planets.  

Detections of debris disks in Band 1 will be challenging compared to detecting
forming condensations in protoplanetary disks.  Debris disks  typically have
relatively low surface brightnesses and large spatial distributions 100s of AU
in radii.  They also can be found much closer to the Sun than the nearest
protoplanetary disks.  Indeed, the closest disks could subtend as much as
$\sim$150$^{\prime\prime}$ on the sky (assuming a 300 AU diameter disk at 2 pc).
Therefore, ALMA's large field-of-view relative to other long wavelength instruments,
such as the JVLA, will be very advantageous for imaging these objects.  (Mosaicking
will still be required to image the largest ones on the sky.)  In addition, the ALMA 12-m 
Array's smaller minimum baselines and the ACA will provide higher sensitivity
to the low surface brightness emission from these objects.

\subsection{The First Generation of Galaxies: \\ Molecular gas in galaxies during the era of re-ionization}
\label{sec:red}

The first generation of luminous objects in the Universe began the process of re-ionizing  the intergalactic 
medium (IGM).  The detection of large-scale polarization in the cosmic microwave background (CMB), caused 
by Thomson scattering of the CMB by the IGM during re-ionization, suggests that the Universe was significantly 
ionized as far back as $z~\approx$~11.0~$\pm$~1.4 (Dunkley et al.~2009).  The ``near" edge of the era of
re-ionization has been inferred from the detection of the Gunn-Peterson effect (Gunn \& Peterson 1965) toward 
galaxies with $z \gtrsim 6$ (Fan et al.\ 2006a,b).  The nearly complete absorption of all continuum shortward of 
the Ly$\alpha$ break is due to moderate amounts of neutral hydrogen in the IGM, suggesting re-ionization was 
complete by $z$ $\approx$ 6.  The Gunn-Peterson effect also insures that at these redshifts the Universe is 
opaque at wavelengths shorter than $\sim$ 1$\,\mu$m.

\begin{figure} 
\includegraphics[scale=0.40]{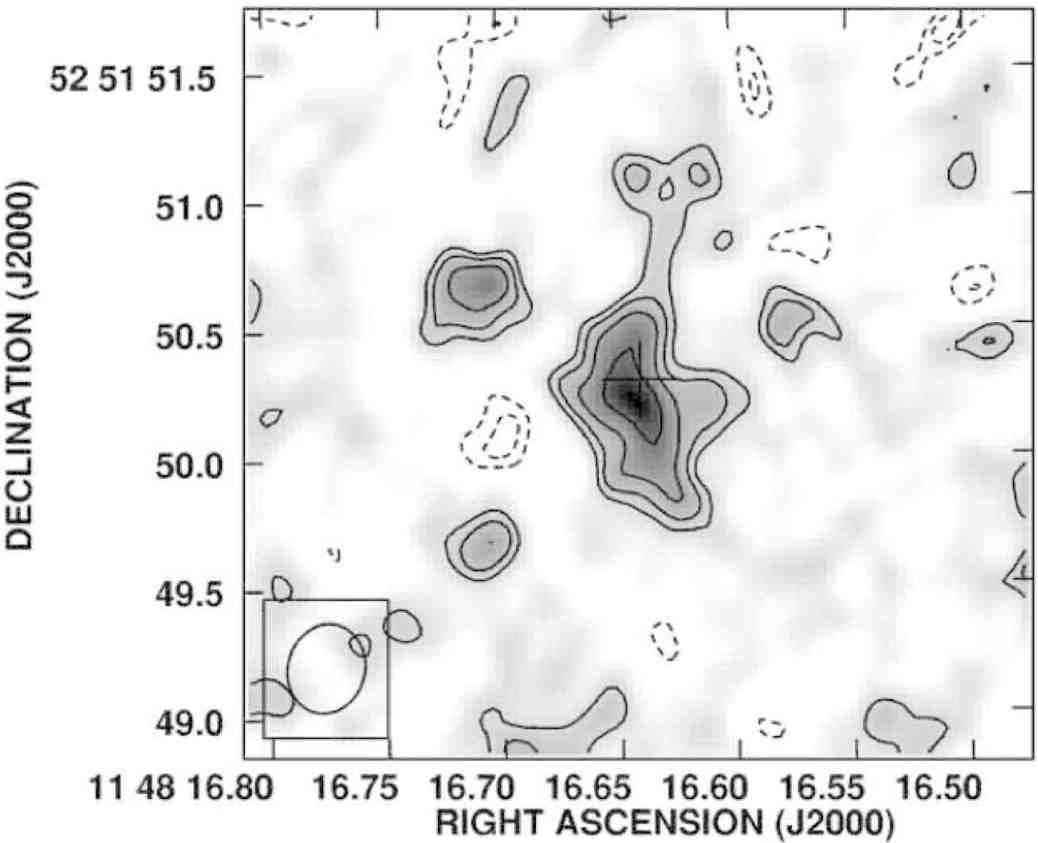}
\caption{VLA redshifted CO $J$=3--2 map of the quasar J1148+5251 using the combined
B- and C-array data sets (covering the 
total bandwidth, 37.5 MHz or 240 km s$^{-1}$), from Walter et al.\ (2004). Contours are 
shown at --2, --1.4, 1.4, 2, 2.8, and 4 $\times \sigma$ (1 $\sigma$ = 43 $\mu$Jy 
beam$^{-1}$). The beam size (0.35\arcsec $\times$0.30\arcsec) is shown in the bottom left corner; the 
plus sign indicates the SDSS position (and positional accuracy) of J1148+5251.}
\label{fig:red1}
\end{figure} 

To study the first generations of galaxies, and to understand the origins of the black hole-bulge mass relation, it will be necessary to study the star-formation properties of galaxies in the $6 \lesssim z \lesssim 11$ range.  Quasar hosts and other sources are rapidly being discovered at the near end of this range (e.g., Cool et al.~2006;  Mortlock et al.~2008; Glikman et al.~2008; Willott et al.~2009), and searches are underway for even more distant objects (e.g., Ota et al.~2008; Bouwens et al.~2009).

Recently, CO has been detected\footnote{Note that interferometers in general have 
an advantage over single-dish telescopes when detecting molecular emission at high redshift since their 
high-resolution imaging capabilities provide the spatial information needed to associate a detection with a 
specific 
object.} in galaxies at redshifts $>$6.  These and other observations in the cm/mm of $z>6$ galaxies are summarized by  Carilli et al.~(2008; 
see also the large surveys of CO at $z$ $>$ 6 by Wang et al.\ 2010, 2011a and references therein).  Current 
instrumentation sensitivities are such that detections are limited to hyperluminous infrared galaxies, i.e. L$_{\rm 
FIR} > 10^{13}$ L$_\odot$.  Only a small fraction of galaxies are this luminous.  The best-studied such object is 
J1148+5251 with a redshift of $z=6.419$ (see Carilli et al.~2008).  For example, Walter et al.\ (2004) imaged the 
CO $J$=3--2 emission (Figure~\ref{fig:red1}) using the VLA, from which they were able to infer the dynamical 
mass.  Walter et al.~(2009) were not able to detect the [NII] line at  205 $\mu$m, but did detect the CO $J$=6--5 
transition.  More recently, Wang et al.\ (2011b) detected the lower-energy CO $J$=2--1 transition and Reichers 
et al.\ (2009) imaged CO $J$=7--6 and CI ($^{3}P_{2}$--$^{3}P_{1}$) emission towards this source.  These and 
other (dust continuum) observations show that there was already a significant abundance of metals and dust by 
this epoch.

\begin{figure} 
\includegraphics[scale=0.60]{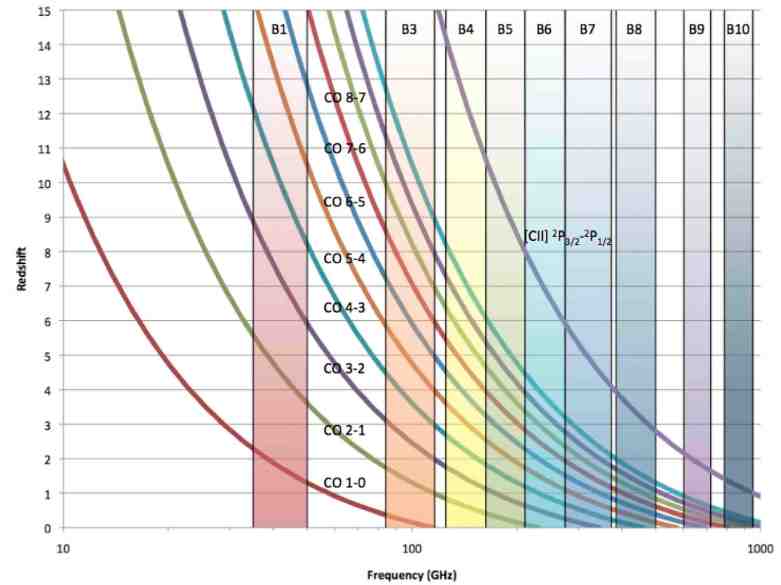}
\caption{Observable frequencies of $^{12}$CO rotational transitions and [CII] $^{2}$P$_{3/2}$--$^{2}$P$_{1/2}$
as a function of redshift.  The frequency ranges of the ALMA Bands are also shown.  Note that the range
for Band 1 reflects the new nominal range of 35-50 GHz.}
\label{fig:red2}
\end{figure}

Figure~\ref{fig:red2} shows the observable frequency of rotational transitions of $^{12}$CO, from $J$=1--0 through $J
$=10--9, as a function of redshift.  Also shown are the frequency ranges of the ALMA Bands (excluding Band 2 for clarity).  
Note that this Figure shows the new nominal range of Band 1 of 35-50 GHz, as this range will yield the highest sensitivities.  
As the Figure shows, Band 1 receivers will be able to detect galaxies in $J$=3--2 at $6 \lesssim z \lesssim 9$, i.e., in the 
redshifts of the era of re-ionization ($z \ga 6$), while higher Bands can only observe higher-$J$ lines that may be less 
excited.  (For example, Band 3 receivers would be able to detect $J$=6--5 emission in the range $4.8 \lesssim z \lesssim 
7.2$.)  Moreover, Band 1 receivers will enable coverage for $J$=2--1 and $J$=1--0 emission at $3.6 \lesssim z \lesssim 
5.6$ and $1.3 \lesssim z \lesssim 2.3$, respectively.  Assuming a 150 $\mu$Jy CO $J$=2--1 line of width
$\sim$600 km~s$^{-1}$ at $z = 5.7$, a 5 $\sigma$ detection would take less than 4 hours with the 50-antenna
ALMA 12-m Array.
 
Band 1 will also allow multiline observations toward certain subsets of redshifts.  For example, galaxies at $1.3 \lesssim z 
\lesssim 2.3$ can be observed in Band 1 but also at $J$=4--3 and $J$=3--2 in Band 4 (NB: a small gap exists at $z$ $
\approx$ 1.8).  Figure~\ref{fig:red2} also shows that in addition the [CII] $^{2}$P$_{3/2}$--$^{2}$P$_{1/2}$ line can be 
observed toward a subset of these galaxies at $1.6 \lesssim z \lesssim 2.2$ using Band 9.  The [CII] line can also be
observed toward galaxies at $2.8 \lesssim z \lesssim 5.9$ using Bands 7 and 8 (NB: a small gap in redshift coverage
exists at $z$ $\approx$ 4).

As with other ALMA Bands, high-redshift science will be done with Band 1 in a targeted mode, i.e., towards known
high-$z$ sources.  An instantaneous $\sim$8 GHz range of frequency coverage, however, will allow significant
sensitivity to other sources proximate on the sky to the known target source but at quite different redshifts.  (If the
target sources are within clustered environments, other sources may even be found at similar redshifts.)  Indeed,
``blank-sky" surveys, made by pointing ALMA towards one location but stepping through the entire Band 1 frequency
range, are an enticing possibility (see, e.g., Aravena et al.\ 2012).  In particular, the ALMA 12-m Array's antennas 
provide a much larger instantaneous field-of-view than the JVLA's antennas, allowing wider searches of blank sky.

In summary, ALMA Band 1 will allow for wide-band observations of molecular emission from many interesting
galaxies in the era of re-ionization.  Band 1 allows for observations of lower-$J$\ lines that are complementary
to lines detected with higher frequency bands.  In particular, ALMA's southern location will allow observations
of objects not observable (well or at all) with the JVLA.  Also, its larger field-of-view gives it an edge in areal 
``blank sky" coverage for detecting at similar or different redshifts sources proximate to known targets.

\subsubsection{Quasar Host Galaxies}

The discovery of molecular gas in quasar host galaxies at $z \sim 6$, when the Universe
 was less than 1~Gyr old 
(Walter et al.\ 2003; Bertoldi et al.\ 2003; Carilli et al.\ 2007),
has opened a new window on the study of gas in systems that
contributed to the re-ionization of the Universe. Studies 
of how the molecular gas properties should evolve,
 and how they can be used to reveal the dynamics of these 
 \textit{massive} systems, have recently prompted a new generation of  
 semi-analytic models with the further aim of understanding how 
 high-redshift quasars fit within the context of large-scale structure
formation. 
Li et al.\ (2007, 2008) have used state of the art
 N-body simulations to show that the observed optical properties of high-redshift
 quasars can be explained if these objects formed early on in the most
 massive dark matter halos ($\sim
 8\times10^{12}$~M$_{\odot}$). These models predict that the 
 most luminous quasars should evolve due to an increase of major mergers, 
which one would expect to find evidence for in the 
CO line profiles and the spatial distribution of the molecular gas
(Narayanan et al.\ 2008). 
Detailed radiative transfer models of the FIR spectral energy distribution of these
systems have been driven by the observations of one 
$z = 6.42$ quasar (namely J1148+5251; Walter et al.\ 2003, 2004).
Larger samples of CO-detected quasars are needed to 
provide better constraints on the models and constrain
dynamical masses to compare with infrared measurements
of black-hole masses (e.g., from MgII lines) and explore the
(possible) evolution of the relation between the masses of 
central black holes and bulges.  Current 3~mm surveys of high-$J$\ CO line
emission in $z\sim 6$ FIR-luminous quasars are being conducted
with the PdBI, having successfully detected CO line emission in
eight objects (Wang et al.\ 2010, 2011a).   Lower-$J$ lines, like 
those accessible with ALMA Band 1, will trace the more abundant
lower density gas in these systems.  Here again, ALMA's 
southern location will prove to be an advantage for targets too 
far south to be well observed with the JVLA.
  
\subsubsection{Lyman-$\alpha$ Emitters}

The rarity of the
luminous quasars at early times suggests that their UV emission was 
unlikely to have contributed
significantly to the re-ionization of the Universe (e.g., Fan et al.\
2001). A more important type of galaxy in the context of cosmic re-ionization
are the Lyman-$\alpha$ emitters (hereafter LAEs).  These galaxies were
discovered through their excess emission in narrow-band images centered 
on the redshifted Lyman-$\alpha$ line  (e.g. Hu et al.\ 1998; Rhoads et al.\ 2000;
Taniguchi et al.\ 2005), and constitute a significant fraction of
 the star-forming galaxy population at $z \sim 6$.
 While the star-formation rates in LAEs inferred from their UV
 continuum emission are a few tens of solar masses per year
 (e.g., Taniguchi et al.\ 2005),
 their number density and the shape of the Lyman-$\alpha$ 
emission line provide important probes of physical 
conditions in the Universe around the epoch of re-ionization. As such,
 it is very important that we understand the properties related to their
 star-formation activity. In particular, we need to quantify the amount of
 molecular gas available for
 fuel. Wagg, Kanekar \& Carilli (2009) used the Green Bank Telescope to
 search for CO $J$=1-0 line emission in two $z > 6.5$ LAEs, including
 the highest spectroscopically confirmed redshift LAE at $z = 6.96$ (Iye et al.\
 2006). The limits to the CO line luminosity implied by the
 non-detections of \co in these two objects suggest modest
 molecular gas masses ($\la$\,10$^{10}$~M$_{\odot}$). 
This conclusion, however, is based on observations of only two
 objects, and future studies would benefit from the sensitivity gained
 by observing higher-$J$\ CO transitions, whose
 flux density may scale as $\nu^2$ due to a contribution to the
 molecular gas excitation by the cosmic microwave background radiation
 (19~K at $z = 6$).   With other facilities, it has been proven challenging to
 detect even the higher energy CO $J$=2--1 line from Lyman-$\alpha$-emitting
 galaxies at these redshifts, using existing facilities (Wagg \& Kanekar 2011).
 At these redshifts, such studies would require ALMA, including the Band
 1 receivers.  Again, ALMA's southern location is advantageous for the 
 detection of more southern LAEs.

\newpage
\section{Suitability of Band 1 for ALMA vs. JVLA}
\label{sec:suitability}

Here we compare the relative capabilities of ALMA and the Jansky Very Large Array
(JVLA) over Band 1 frequencies in common.  The JVLA currently has observing capability
over the nominal Band 1 frequency range of 35--50 GHz, through its receivers in the
K$_{\rm a}$-band (26.5--40 GHz) and Q-band (40--50 GHz). ALMA Band 1, however,
will likely be extended to 50-52 GHz, frequencies the JVLA cannot observe.   In the
following, we compare the differences in site conditions and array characteristics that
show that Band 1 observing is superior with ALMA than with the JVLA.

\subsection{Site Conditions}

ALMA is located on the Llano de Chajnantor at a higher altitude (5040 m) than the
JVLA on the Plains of San Agustin (2124 m).   Opacity in Band 1 consists of a wet
component of atmospheric water vapor and a dry component of non-H$_{2}$O gases,
like O$_{2}$.  The quantity of the wet component, as measured by precipitable water vapor
(PWV) affects more the lower end of the Band 1 frequency range.  The dry component,
however, dominates at the upper end.  Nevertheless, the ALMA site is very well-suited
for Band 1 observing.  Even during the worst octile of weather, however, the typical
optical depth through the Band 1 Receiver range is less than 0.1.  Though other
frequency ranges like Band 3 can still use such weather, the addition of cloud cover
and water droplets in the air make even lower frequency observations more attractive. 

The PWV over the JVLA during the years 1990-1998 was measured to range between
4.5 mm in winter and 14 mm in summer with a $\pm$2 mm scatter throughout the year
(Butler 1998; VLA Memo 237).  In comparison, the PWV over ALMA during the years
1995-2003 was measured to range between 1.2 mm in winter to 3.5-7.0 mm in summer
(median $\approx$ 1.4 mm), using opacity data obtained by Ot\'orola et al. (2005; ALMA
Memo 512) and conversions provided by D'Addario \& Holdaway (2003; ALMA Memo
521).  For frequencies $<$45 GHz, Butler (2010; VLA Test Memo 232) found empirically
a linear relation between opacity and PWV, where opacities varied from 6\% to 10\%
from 1 mm to 14 mm.  Assuming this relation is applicable to both observatories, we
find the atmospheric opacities at $<$45 GHz over ALMA to be generally half those
over JVLA.

Phase stability over the JVLA was measured with a 300-m baseline test interferometer at
11.3 GHz, and median characteristics from one year of data were reported by Butler \&
Desai (1999; VLA Test Memo 222).  They found median phase variation rms values ranging
from 2-2.5$^{\circ}$ in winter nighttime to $>$10$^{\circ}$ in summer daytime.  Scaling
these values to the zenith and converting to path delay rms fluctuations, these phases
convert to 430-540 fsec to 2100 fsec, respectively.  For ALMA, D'Addario \& Holdaway,
using six years of data from a similar 300-m baseline test interferometer, determined a
median path delay fluctuation of 500 fsec.  (A seasonal breakdown was not provided.)
Though the data are somewhat scant, the overall median path delay at ALMA is about
equal that of the best median path delays at the JVLA in winter nighttime.  Note, however,
that phase stability can be mitigated by water vapor radiometer data available at both
sites.

\subsection{Array Characteristics}


The most important difference between the characteristics of ALMA and the JVLA is that they are located at
very different latitudes,  the former at $-23\degree$ and the latter at $+34\degree$.  Some sources too far 
south to be observed at the JVLA (or at least observed well) will be observable with ALMA.  (The Australia
Telescope Compact Array (ATCA) can also observe some Band 1 frequencies from the southern hemisphere
but at much lower relative sensitivity than ALMA or the JVLA.  Hence, we do not consider it further.)  Important
targets in the southern hemisphere that are better observed at ALMA than the JVLA (if at all) include Sgr~A*,
the center of our Galaxy, the Magellanic Clouds, the closest neighboring galaxies, and TW Hya, the closest
protoplanetary disk.  Indeed, any source observed with ALMA in higher frequency bands can be more
effectively observed at 35-50 GHz with Band 1 receivers.  Also, with numerous satellite observatories providing
full sky coverage (e.g., JWST, {\it Spitzer}, {\it Herschel}\/), having full-sky coverage from ground-based facilities
at important frequencies is optimal.

\begin{table*}[h]
  \caption{Summary of general properties of the ALMA Band 1 and JVLA}
\label{table:general}
  \begin{center}
    \begin{tabular}{lcc}
      \hline
      & ALMA \\
      & Band 1 & JVLA \\
      \hline
      \noalign{\smallskip}
      Latitude  & $-23\arcdeg$ & $+34\arcdeg$ \\
      Altitude {\small (m)} & 5040 & 2124 \\
      No. of antennas & 50 & 25 \\
      Antenna diameter & 12 & 25 \\
      Pointing accuracy {\small (arcsec)}& 0.6 & 2--3\\
      Frequencies {\small (GHz)} & 35--52 & 26.5--40 (band K$_a$)\\
      &         & 40--50 (band Q)\\
      Aperture efficiency, $A_{\rm e}$ & 0.78 & 0.34--0.39 \\
      $\Delta\nu_{\rm max}$ (Hz) & 3820 & 1 \\
      \hline \noalign{\smallskip}
      Single-field sensitivity ($\propto ND^2$) & 7200 & 17000 \\
      \textit{effective} & 5600 & 5800-6600 \\
      \noalign{\medskip}
      Mosaic image sensitivity ($\propto ND$) & 600 & 680 \\
      \textit{effective} & 530 & 420--390 \\
      \noalign{\medskip}
      Image fidelity ($\propto N^3$) & 130000 & 20000\\
      \noalign{\smallskip}
      \hline
      
      \hline

    \end{tabular}
  \end{center}
\end{table*}

Table 1 summarizes the differences between ALMA and the JVLA.
Comparing their attributes, we note that ALMA's 12-m Array has antennas of smaller
surface area than those of the JVLA (12-m diameter vs.\ 25-m) but these are larger in number (50 in the 12-m
Array vs.\/~27) and have higher pointing accuracies (0.6\arcsec\ vs.\ 2-3\arcsec) and aperture efficiencies at
Band 1 frequencies (0.78 vs.\ 0.34--0.39).  Combining these numbers (except pointing accuracy), the effective
surface area of the ALMA 12-m Array is a factor of 0.85--0.98 that of the JVLA.  Adding Band 1 to the ACA
antennas would minimize even this small difference.  ALMA has the same 8~GHz maximum bandwidth as the
JVLA with its new WIDAR correlator.  ALMA's present correlator has a lower maximum spectral resolution than
the JVLA's, however, i.e.,  a maximum of 3.82~kHz~vs.~1~Hz, respectively.  (ALMA's correlator of course could
be similarly upgraded in the future.)
 
\begin{figure}
\begin{center}
\includegraphics[scale=0.55]{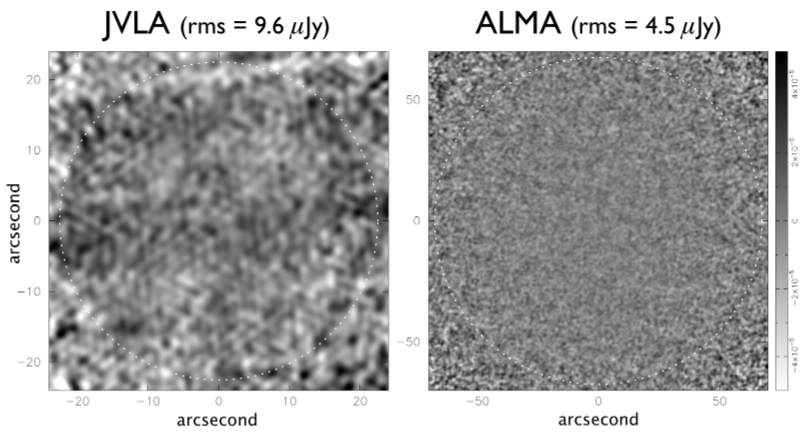}
\caption{Images from JVLA and ALMA observations simulated with CASA. 
The observations were set toward a ``blank" sky at 45~GHz with 8~GHz (continuum)
bandwidth, with JVLA in its D-configuration while ALMA in its ``12" configuration
provided in CASA.  Both array configurations give rise to a similar angular resolution
of $\sim$1\farcs6 FWHM.  The white dotted circles denote the corresponding primary
beam sizes. There resulting 1 $\sigma$ rms noise levels after 2 hours of on-source
integration are 9.6 $\mu$Jy and 4.5 $\mu$Jy, respectively, for JVLA and ALMA, 
which are in general agreement with the estimated noise level shown in
Table~\ref{table:sensi_comp}.}
\label{fig:blank}
\end{center}
\end{figure} 

 \begin{table*}[h]

\caption{Comparison of Point-Source Sensitivity between JVLA and ALMA}
\label{table:sensi_comp}
\begin{center}
\begin{tabular}{cccccc}
\hline
   \multicolumn{2}{c}{}                        & \multicolumn{2}{c}{JVLA}      & \multicolumn{2}{c}{ALMA} \\
\hline
   \multicolumn{2}{c}{no. of antennas}& \multicolumn{2}{c}{25}         & \multicolumn{2}{c}{50} \\
   \multicolumn{2}{c}{polarization}    & \multicolumn{2}{c}{dual}              & \multicolumn{2}{c}{dual} \\
   \multicolumn{2}{c}{weather}         & \multicolumn{2}{c}{winter}    & \multicolumn{2}{c}{auto (5.2 mm) PWV} \\
   \multicolumn{2}{c}{source position} & \multicolumn{2}{c}{zenith}    & \multicolumn{2}{c}{zenith} \\
\hline
\hline
   \multicolumn{2}{c}{on-source time}  & 60~s & 1~hr & 60~s & 1~hr \\
\hline
   \multicolumn{2}{c}{bandwidth}       & \multicolumn{2}{c}{1 MHz}      & \multicolumn{2}{c}{1 MHz} \\
\hline
   freq. & 35 GHz   & 3.2~mJy   &  0.41~mJy   &  3.0 mJy  & 0.38 mJy   \\
            & 40 GHz   & 3.6~mJy   &  0.47~mJy   &  3.1 mJy  & 0.40 mJy   \\
            & 45 GHz   & 5.1~mJy   &  0.68~mJy   &  3.6 mJy  & 0.47 mJy   \\
            & 50 GHz   & 25.5~mJy   &  3.29~mJy   & \multicolumn{2}{c}{(not available)}   \\
\hline
   \multicolumn{2}{c}{bandwidth}               & \multicolumn{2}{c}{8 GHz}      & \multicolumn{2}{c}{8 GHz} \\
\hline
    freq. & 40 GHz   & 50~$\mu$Jy   &  5.4~$\mu$Jy   & 35~$\mu$Jy  & 4.5~$\mu$Jy   \\
             & 45 GHz   & 78~$\mu$Jy   &  10~$\mu$Jy   & 41~$\mu$Jy  & 5.3~$\mu$Jy   \\
\hline
\end{tabular}
\end{center}
\end{table*}

Given differing antenna numbers, sizes, and baselines, the two observatories differ in various imaging
metrics{\footnote{These metrics were defined and used to compare ALMA to other existing interferometers
in the 2005 NRC document {\it The Atacama Large Millimetre Array: Implications of a Potential Descope}.}:

\begin{figure}
\begin{center}
\includegraphics[scale=0.65]{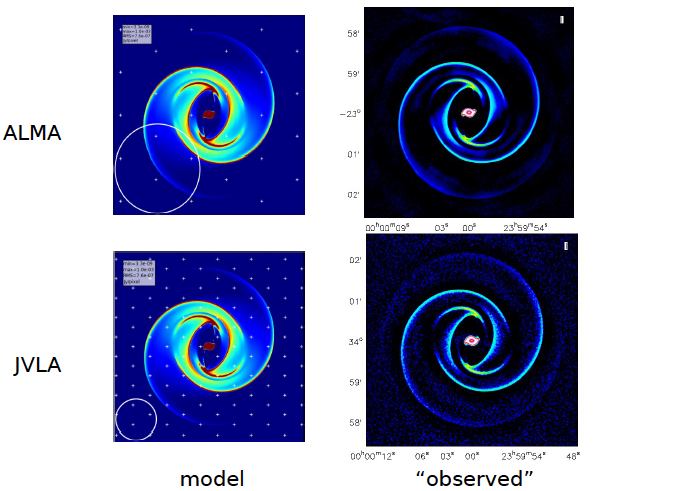}
\caption{
Images from CASA simulations of JVLA and ALMA mosaic observations of 45 GHz continuum. The left-hand panels show the model image convolved with the synthesized beams. The pointing patterns for the mosaicked observations are shown with white dots. The right-hand panels show the resulting observed images.  Both simulated observations are executed with eight hours of 
on-source time in total toward the zenith. The ALMA and JVLA were assumed to be in their ``12" and ``D" configurations, (both provided in CASA), respectively, which resulted in similar synthesized beam sizes of 1.7$^{\prime\prime}$ $\times$ 1.7$^{\prime\prime}$. The achieved noise level by ALMA is around three times better than that by JVLA. (i.e,. 10 $\mu$Jy~beam$^{-1}$ for ALMA vs. 30 $\mu$Jy~beam$^{-1}$ for JVLA).  Observation overheads (e.g., calibration scans) and phase decoherence due to site location were not included in the simulations, both of which will lead to greater degradation
in the  JVLA images.
}
\label{fig:mosaic}
\end{center}
\end{figure} 

\begin{figure}
\begin{center}
\includegraphics[scale=0.65]{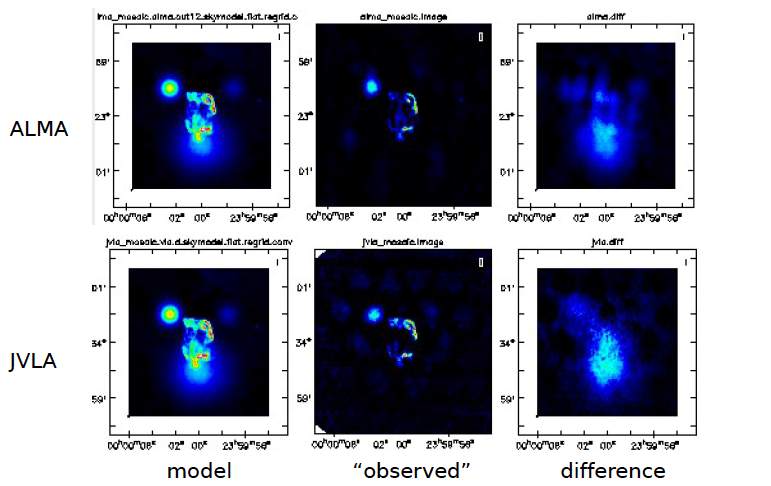}
\caption{
Images from CASA simulations of observations of extended 45 GHz emission with the JVLA and ALMA. The left-hand panels show the model image (a superposition of the G41.1-0.3.b template provided by the CASA guide with three extended Gaussian sources (two 18$^{\prime\prime}$ in size and one 48$^{\prime\prime}$ in size) convolved with the synthesized beams. The middle panels show the resulting images from the simulations. The right-hand panels show the difference between the model and observation images.  Both simulated observations were executed with one hour of on-source time in total toward the zenith. The ALMA and JVLA are assumed to be in their ``12" and ``D" configurations (both provided in CASA), respectively, which resulted in similar synthesized beam sizes of 1.7$^{\prime\prime}$ $\times$ 1.7$^{\prime\prime}$. The achieved noise level by ALMA is around five times better than that by JVLA (i.e., 10 $\mu$Jy~beam$^{-1}$ for ALMA vs. 50 $\mu$Jy beam$^{-1}$ for JVLA). Observation overheads (e.g., calibration scans) and phase decoherence due to site location were not included in the simulations, both of which will lead to greater degradation in the JVLA images.
}
\label{fig:mosaic_fidelity}
\end{center}
\end{figure} 
\begin{itemize}
	\item Comparing the face-value ``single-field sensitivity" metric ($ND^{2}$; where $D$ is the antenna diameter and
          $N$ is the number of antennas), ALMA appears about half as ``sensitive" as the JVLA (7200 vs.~17000).  Factoring
          in aperture efficiencies to give effective values of $D$, however, the metrics are actually much more similar
          (5600 vs.~5800-6600).  Table~\ref{table:sensi_comp} provides more realistic comparisons of JVLA and ALMA
          sensitivities for point sources across the proposed Band~1 frequency range, estimated using their respective
          sensitivity calculators\footnote{For JVLA and ALMA, see {\it
              https://science.nrao.edu/facilities/evla/calibration-and-tools/exposure}\/ and {\it
              http://almascience.eso.org/call-for-proposals/sensitivity-calculator}, respectively.  For these calculations,
            we assume the original ALMA specifications for Band 1 receiver performance, i.e., the same 40--80 K as for the
            JVLA's K$_{\rm a}$/Q-band receivers.}.  Note that the JVLA sensitivities require the JVLA's best weather
          (``winter") while a relatively high PWV level (5.2 mm) was actually chosen for ALMA here.  From these
          calculations, we see continuum sensitivities of ALMA for Band 1 are actually {\it similar to better}\/ than those
          of the JVLA.  For example, a 1 $\sigma$ rms of $\sim$5 $\mu$Jy beam$^{-1}$ is expected at 40 GHz after 1 hour of
          integration at both observatories.  At higher frequencies (e.g., $>$45~GHz), however, the point source sensitivity
          of ALMA is {\it better}\/ than that of JVLA by factors of 1.4--1.9, depending on bandwidth.  (Simulations of JVLA 
          and ALMA observations suggest even larger improvements; see below.)  In addition,
          Figure~\ref{fig:blank} shows simulated ``blank-sky" observations at 45 GHz carried out with CASA, giving another
          perspective on this comparison.  (ALMA's improved pointing accuracy and better phase stability were not fully
          incorporated into these calculations.)  Note also that ALMA's 12-m diameter antennas provide a field-of-view for
          single-pointing observations that is more than twice as wide as what the JVLA's 25-m diameter antennas provide
          (see Table~\ref{table:beam_comp}), so ALMA's similar or better sensitivity is obtained over a wider area in a
          single pointing.

	\item Comparing the ``mosaic image sensitivity" metric ($ND$), again on face value, ALMA's 12-m Array and the JVLA
          appear already quite similar (600 vs.~680, respectively).  Factoring in only the improved aperture efficiencies of
          ALMA at its lowest frequencies vs.~those of the JVLA at its highest frequencies, the comparison is in ALMA's
          favour by a factor of $\sim$1.3 (530 vs.~420-390).  As with the single-pointing comparison above, the superior
          weather at the ALMA site will increase this factor further.  For example, Figure~\ref{fig:mosaic} shows mosaic
          simulations for JVLA and ALMA of a galaxy, in relatively similar compact configurations over the same 8 hours 
          of integration. The ALMA observations are performed with fewer pointings than those of the JVLA.  The
          resulting 1 $\sigma$ rms noise level of the ALMA image is a factor of three better than that of the JVLA image.
	
	\item ALMA's larger number of baselines yield a higher ``image fidelity" metric ($N^3$) by a factor of $>$6
	(130000 for ALMA vs.\ 20000 for the JVLA) over similar observation durations.  Basically, ALMA's larger
	number of baselines allow more spatial frequencies to be sampled per unit time, yielding more accurate
	images.  Figure~\ref{fig:mosaic_fidelity} shows an example of ALMA's higher intrinsic fidelity relative to 
	that of the JVLA, especially for extended emission, based on simulations of a high-mass star-forming region.
	The difference between the model and observation images (right panel) is noticeably smaller for the ALMA
	case than for the JVLA one.
	
\begin{table*}[h]

\caption{Comparison of angular scale coverage between JVLA and ALMA at 45 GHz}
\label{table:beam_comp}
\begin{center}
\begin{tabular}{ccccc}
\hline
			& \multicolumn{2}{c}{JVLA}	& \multicolumn{2}{c}{ALMA} \\
\hline
 Configuration	& A & D & most extended & most compact \\
 B$_{min}$ (km)	& 0.68	& 0.035	& 0.04	& 0.015	\\
 B$_{max}$ (km) 	& 36.4	& 1.03	& 16		& 0.15 	\\
 $\theta_{PRIMARY}$& 60	& 60		& 135	& 135	\\
 $\theta_{HFBW}$	& 0.043	& 1.5		& 0.08	& 9		\\
 $\theta_{LAS}$	& 1.2		& 32		& 35		& 93		\\
\hline
\end{tabular}
\end{center}
\end{table*}

	\item At present, ALMA has maximum baselines that are a factor of $\sim$2 smaller than the
	JVLA's (15-18~km vs.~36.4~km), meaning that the JVLA can in principle produce images of resolution
	up to a factor of 2 higher than ALMA can at the same frequency.  ALMA will be in turn more sensitive to
	extended emission, however.  First, ALMA's smaller dishes mean that its minimum baselines are shorter
	than those of the JVLA (16-m~vs.~35-m; see Table~\ref{table:beam_comp} for a comparison), allowing
	higher sensitivity to extended, low-surface-brightness emission.  Second, ALMA can include the ACA
	antennas, each of 7-m diameter but together in a close-packed configuration, in principle allowing even
	further sensitivity to extended emission.
	
\end{itemize}

In summary, ALMA Band 1 can be superior to the JVLA at its highest frequencies in many ways, including:

\begin{itemize}

\item {\bf Access to southern sources}, given ALMA's southern hemisphere location; 

\item {\bf Wide-field sensitive imaging}, due to ALMA's larger number of smaller, high precision antennas
located at an excellent site; 

\item {\bf High image fidelity}, given ALMA's larger number of antennas; 

\item {\bf Sensitivity to extended emission}, if appropriate, due to ALMA's shorter minimum baselines and the
ACA; 

\item {\bf Likely coverage of 50-52 GHz}, frequencies not possible with the current JVLA receivers;

\item {\bf Recovery of short spacing visibilities}, by using the Atacama
  Compact Array, and the total power single-dish observations;

\item {\bf Combination with other ALMA bands}, for many multi-band projects; and

\item {\bf Lower overheads}, by applying for and using a single observatory.

\end{itemize}

As shown in \S 4, the top science cases for Band 1 can stand shoulder-to-shoulder with the primary
Level 0 goals of ALMA.  Thus, the primary motivation for the enhancement is {\it not}\/ as a ``poor weather"
back-up receiver but rather the excellent science that can be achieved.  In the following sections, we 
explore the large and broad variety of science cases beyond the top cases identified in \S4 that the
ALMA Band 1receiver suite will be able to address.

\newpage
\section{A Broad Range of Science Cases}
\label{sec:range}

Along with the two science cases presented above in \S4, there is a wealth of scientific opportunity available to the wide 
ALMA community when the Band 1 receiver suite is built.  Here we highlight a selection of science cases which would 
significantly benefit from Band 1 receivers on ALMA.

\subsection{Continuum Observations with ALMA Band 1}

The astrophysical continuum radiation at wavelengths of $\sim$1\,cm is
relatively unexplored. Yet, this radiation is key to understanding radio
emission mechanisms and probing regions that are optically thick at
shorter wavelengths.  The sensitivity and resolution of ALMA Band 1
will allow: (1) improved understanding of galaxy clusters through the 
Sunyaev-Zel'dovich Effect; (2) a diagnostic of the smallest interstellar
dust grains; (3) studies of jets from young stars; (4) an understanding of
the nature of pulsar wind nebulae; (5) the detection of radio SNe, with
constraints on stellar precursors and remnants; (6) a diagnostic of X-ray
binaries; and (7) improved probes of Sgr~A*, the supermassive black
hole at the center of the Galaxy.

\subsubsection{The Sunyaev-Zel'dovich Effect}

Much of what we know about galaxy clusters has come from X-ray
observations of thermal bremsstrahlung emission of the intra-cluster
medium (ICM).  For example, the angular resolution of {\it Chandra}
has been crucial to advancing our understanding in this area and has
resulted in a renaissance in astrophysical studies of galaxy
clusters.  In recent years, the Sunyaev-Zel'dovich Effect (SZE) has
provided an increasingly important view of these cosmic structures
(Birkinshaw 1999).  Since the SZE signal is proportional to the
product of the electron density and its temperature ($\sim n_e \,T_e$,
compared to $n_e^2\sqrt{T_e}$ for the X-rays), it gives a complementary
view of the physical state of the ICM, one more sensitive to hot phases
that also directly measures local departures from thermal pressure
equilibrium.  To date, the majority of SZE observations have been
carried out at comparatively low angular resolution (beams $> 1'$ in
size), yielding information about the overall bulk cluster properties.
Advances in instrumentation have begun making higher angular
resolution measurements of the SZE possible, revealing previously
unsuspected shock-heated gas in the ICM of clusters previously
thought to be dynamically relaxed (Komatsu et al.\ 2001, Kitayama
et al.\ 2004, Mason et al.\ 2010, Korngut et
al.\ 2011, Plagge et al.\ 2012).  These $10''$ to $20''$ SZE images are
the current state of the art.  A Band 1 receiver suite on ALMA will
surpass this benchmark, making possible detailed studies of the ICM
using the SZE on larger samples and with greater sensitivity than before.

ALMA Band 1 receivers will be capable of addressing a wide range of
basic questions about the observed structure and evolution of clusters.
For example, what are the structures of ICM shocks and the mechanisms
responsible for converting gravitational potential energy into thermal
energy in the ICM (Markevitch et al.\ 2007, Sarazin et al.\ 1988)?  What
is the influence of Helium ion sedimentation within the cluster atmosphere
(Ettori et al.\ 2006)? What is the nature of the AGN-inflated ``bubbles'' seen
in the cores of some clusters (Pfrommer et al.\ 2005), and what is the role
of cosmic rays in the ICM? What is the nature of the underlying ICM
turbulence (e.g., Kolmogorov versus Kraichnan)? A particularly rich area
will be the detailed study of ICM shocks, which are common since infalling
sub-clusters are typically transsonic.  Several galaxy cluster mergers have
been observed recently with Chandra and XMM in X-rays with resolutions at
the arcsecond level where substructures become visible (Markevitch, et
al.\ 2000, 2002).  The features of interest for these studies will typically fit
within one or a few ALMA Band 1 fields-of-view and require longer integrations
(several to $\sim$10 hours per pointing).  Note that Band~1 receivers also 
may have the sensitivity to detect the SZE from the halos of massive
individual ellipticals or massive groups.

Another important area where high-resolution SZE imaging will have an
impact is the interpretation of SZE survey data. ACT (Dunkley et al.\ 2011),
SPT (Williamson et al.\ 2011), and {\it Planck}\/ (Planck Collaboration, 2011)
have all conducted 1000$+ \, {\rm deg^2}$ surveys to detect and
catalog galaxy clusters via the SZE.  These surveys provide unique and
valuable information about cosmology but their interpretation depends
upon assumptions about the relationship between the SZE signal and
the total virial mass of the halos observed. It is known that both gravitational
(cluster merger) and non-gravitational processes (AGN and supernova
feedback, bulk flows\footnote{By bulk flow, we refer to the motion of a 
cluster itself through its surrounding medium, producing a kinematic
contribution to the observed SZE signal; in theory, this contribution has
a different spectral dependence than the thermal SZE and may be 
distinguishable with good spatial coverage.}, cosmic ray pressure)
give rise to considerable scatter and potential biases (e.g., Morandi
et al.\ 2007) in this relationship. 
Cluster mergers have a particularly dramatic effect on
the SZE, typically generating transsonic (Mach $ \sim 2$-$4$) shock
fronts which can enhance the peak SZE in the cluster by an order of
magnitude (Poole et al.\ 2007, Wik et al.\ 2008).

The systematic astrophysical uncertainties just described are the limiting
factor in making cosmological inferences from the small published samples
of a few dozen SZE-selected clusters (e.g., Sehgal et al.\ 2011).  ALMA Band 1
receivers are the only foreseen prospect for efficient high-resolution observations
of the large southern hemisphere samples of SZE-selected clusters that
will directly improve inferences from these surveys.  They will be used to
image (at $5''-10''$ resolution) galaxy clusters discovered in the
low-resolution ($\sim$1$^{\prime}$) surveys, detecting shocks and mergers and
identifying ICM substructure, and providing a direct, phenomenological
handle on important survey systematics.  Indeed,
the sensitivity and resolution of an ALMA Band 1 receiver suite allows for
efficient follow-up observations of cluster detections made by blind southern
hemisphere SZE surveys. Thus, a study of a selection of clusters from these
survey experiments in a statistical manner becomes feasible and new important
insights into the mass-observable relation and its scatter and dependence
on cluster physics can potentially be obtained. The ability to understand
cluster selection in detail is essential to derive reliable constraints on
cosmological models from SZE cluster surveys (see e.g., Geisbuesch
et al.\ 2005; Geisbuesch \& Hobson 2007).

The coming decade will also see an explosion of optical and X-ray
cluster data.  The German/Russian satellite {\it eRosita}, due
to launch in 2014, will carry out the first all-sky X-ray survey since ROSAT
(Merloni et al.\ 2012).  Among other things, it is expected to catalog
$\sim$100000 clusters out to $z=1.3$ (Cappellutti et al.\ 2011).  Also, the
Dark Energy Survey (DES; Dark Energy Survey Collaboration 2005) is a
$5000 \, {\rm deg^2}$, mostly southern sky survey also expected to find
$\sim$100000 galaxy clusters. Targeted SZE observations with ALMA
Band 1 receivers will be invaluable to determine the properties of clusters
at redshifts where X-ray spectrscopy and gravitational lensing begin to fail.
These high-$z$ clusters, such as the ACT-discovered SZE cluster ``El Gordo''
at $z=0.89$, weighing in at $M=(2.16 \pm 0.32) \times 10^{15}$ M$_{\odot}$
(Menanteau et al.\ 2011), offer leverage on so-called ``pink elephant''
tests capable of constraining cosmological or gravitational theories
based on the existence of individual extreme objects, i.e., provided their
properties are accurately determined.  Importantly, note that an ACA 
equipped with Band 1 receivers will be comparable in capability to the
OVRO/BIMA arrays used in the current decade to measure the bulk
SZE properties of large northern hemisphere cluster samples (Bonamente
et al.\ 2008). Extending this capability to the southern hemisphere over
the next decade is important to realize the full potential of these rich
cluster samples.

Given the large number of ALMA baselines, the resulting high image
fidelity and dynamic range of the data will be advantageous to SZE studies,
in particular the detailed ones.  In addition, long baseline data from ALMA
can be used to  remove accurately the intrinsic and background (i.e.,
gravitationally lensed) discrete source populations. These latter objects
are a signal of substantial interest from another point of view, but they
also set a significant ``confusion noise'' floor to millimeter single-dish
observations, especially considering the factor of $2-3$ boost in source
confusion in clusters due to gravitational lensing (Blain et al.\ 2002).

\begin{figure}
\begin{center}
\includegraphics[scale=0.6]{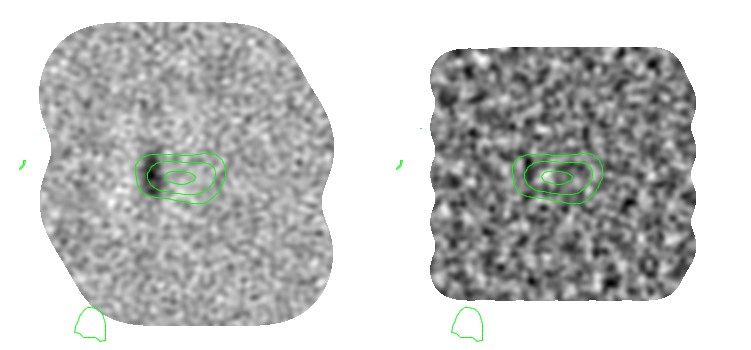}
\caption{Simulated $1.5$~hour ALMA Band 1 (left) and Band 3 (right)
  observations of a galaxy cluster covering $5' \times 5'$. The shock
  is represented as a Gaussian component $5'' \times 25''$ in extent
  with a peak SZE of $y=10^{-4}$, considerably weaker than the
  amplitude observed in RXJ1347-1145 by Mason et al.\ (2010). The Band
  3 data were tapered to the innate resolution of the Band 1 map,
  $\sim 10''$ (FWHM). ACA baselines were not included in this
  simulation but the overplotted contours show the ACA Band 1 image
  (using a $45''$ taper) of the bulk ICM in this system in a simulated
  12 hr integration after subtraction of the shock signal.\ The bulk ICM
  is modeled as an elliptical isothermal $\beta$ model with $R_{core}
  = (150, 250) \, {\rm kpc}$, $\beta = 0.7$, and $y_o = 3 \times
  10^{-5}$ at $z=0.7$, characteristic of disturbed, merging systems.}
\label{fig:szshock}
\end{center}
\end{figure}

ALMA will have a considerably higher sensitivity for these
observations than the JVLA, owing to an order of magnitude
higher surface brightness sensitivity, or ALMA Band 3, owing to
lower system temperatures and larger primary beam.  In Figure
\ref{fig:szshock}, we show simulated Band 1 and Band 3 observations
(using the ALMA 12-m Array and the ACA) that cover the virial region
($D\sim 5'$) of a moderately massive SZE cluster with a merger
shock.  For these simulations, we considered a hypothetical
project to detect a feature with a Compton $y = 10^{-4}$,
characteristic of strong shocks in major mergers, with a characteristic
feature size of $5''-20''$. The required flux density sensitivity is similar
in both cases after allowing for resolution effects, about 1 $\sigma$ =
$8-9 \, {\rm \mu Jy}$ rms  in both instances.  We find that a clear
detection is achieved in only $1.5$ hours of Band 1 observing, but
nearly $40$~hours are required at Band 3.  The ACA Band 1 measurement
of the bulk ICM signature (a 12 hr observation is needed for good SNR)
is also shown, tapered to a $45''$ FWHM beam. Yamada et al.\ (2012)
find similar results in their detailed study of SZE imaging with ALMA
and the ACA at $\lambda \approx 1 \, {\rm cm}$.

In summary, ALMA's southern location matching large galaxy cluster 
surveys, intrinsically high image fidelity, and sensitivity to extended
low-brightness features (e.g., relative to the JVLA) will make Band 1
observations very compelling probes of physics of galaxy clusters
using the Sunyaev-Zel'dovich Effect.

\subsubsection{Very Small Grains and Spinning Dust} 

The last decade has seen the discovery of surprisingly bright cm-wavelength 
radio emission from a number of distinct galactic objects but most notably 
dark clouds (e.g., Finkbeiner et al.\ 2002; Casassus et al.\ 2008 (see 
Figure~\ref{fig:cont1}); Scaife et al.\ 2009).  The spectrum of this new 
component of continuum radiation can be explained by electric dipole radiation 
from rapidly rotating (``spinning") very small dust grains (VSGs), as 
calculated by Draine \& Lazarian (1998; DL98).  This emission has been also
seen as a large-scale foreground in CMB maps, spatially correlated with
thermal dust emission and having a spectrum peaking at $\sim$40 GHz.

All of the existing work aimed at diagnosing this continuum emission is 
derived from CMB experiments on large angular scales, where the bulk
of the radio signal occurs, e.g., recently by the {\it Planck\ }\
satellite.  Details on small angular scales are crucial, however, for probing
star formation and circumstellar environments.  Simply, progress in the 
understanding of the solid and gaseous states of the ISM requires sufficient 
resolution to separate the distinct environments.   Directly measuring the
VSG abundance and solid state physics is very exciting because VSGs play a
central role in the chemical and thermal balance of the ISM.  For example,
the smallest grains account for most of the surface area available for 
catalysis of molecular formation.

\begin{figure}
\epsscale{0.8}
\includegraphics[scale=0.6]{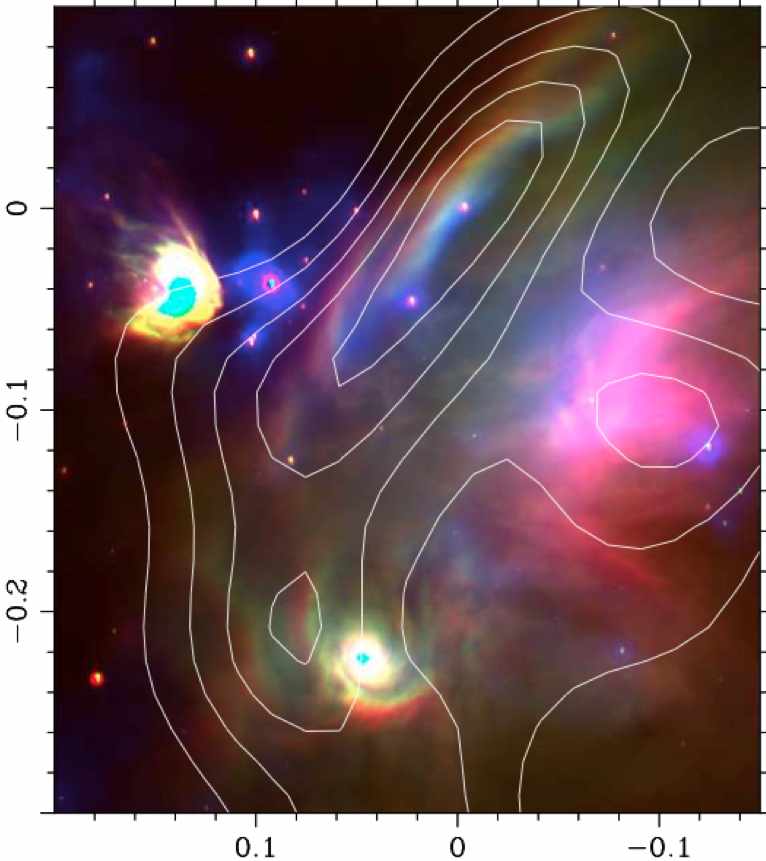}
\caption{Three-colour image of the $\rho$~Oph~W photo-dissociation
  region (Casassus et al.\ 2008). {\bf Red}: MIPS~24~$\mu$m continuum {\bf Green}:
  IRAC 8~$\mu$m continuum,
  dominated by the 7.7~$\mu$m PAH Band {\bf Blue}: 2MASS~K$_{\rm s}$-band
  image. The $x-$ and $y-$axes show offset in RA and Dec from $\rho$~Oph~W,
  in degrees.  The contours follow the 31~GHz emission, with levels at
  0.067, 0.107, 0.140, 0.170, and 0.197  MJy~sr$^{-1}$.   }
\label{fig:cont1}
\end{figure}

DL98 proposed that the grain size distribution in their spinning dust model 
would be dominated by VSGs, thought to be mostly PAH nanoparticles.  The
size distribution of VSGs is poorly known, however, since studies of interstellar
extinction are relatively insensitive to its details.  The existence of VSGs has
been supported by several assertions.  First, a significant amount of carbonaceous 
nanoparticles in the ISM could explain observations of unidentified IR 
emission features.  Second, the strong mid-infrared emission component
seen by IRAS  must result from the reprocessing of starlight by ultrasmall
grains.  Indeed, the fraction of the ISM carbon content proposed to exist in
VSGs considerably exceeds that implied to exist in the MRN dust size
distribution.  (The MRN dust distribution is known to underestimate this
fraction.)  

Observationally determining PAH content in dust clouds is not straightforward.
Where there is a strong source of UV flux present, it is possible to identify 
PAHs by their spectral emission features.  In the case of pre-stellar and Class 
0 cloud cores, however, these features are absent.  With observations
from ALMA Band 1 receivers constraining the spinning dust SED at similar
resolution to, e.g., Spitzer or the forthcoming MIRI instrument on the JWST,
it will be possible to measure the VSG size distribution directly from the data.  

This work will 
also be important in the context of circumstellar and protoplanetary disks, 
where the proposed population of VSGs may have important implications for disk 
evolution.  Certainly, spinning dust emission will provide a better measure of 
the small grain population within circumstellar disks than PAH emission since 
favorable excitation conditions for PAHs exist only in the outermost layers 
of the disk.  Since all the VSGs in the disk should contribute spinning dust 
emission, such emission will provide a much better probe of the mass in VSGs.  
Combining this information with the PAH emission features would then also give 
us a useful measure of sedimentation in disks.

Spinning dust emission from a VSG population will in theory dominate the
thermal emission from disks (around Herbig Ae/Be stars) at frequencies $\leq$
50~GHz by significant factors  (Rafikov 2006).  The existence of these VSGs
has been confirmed observationally from PAH spectral features seen in the 
disks of Herbig Ae/Be stars (Acke \& van den Ancker 2004) but it has not been 
detected in protoplanetary disks due to a lack of strong UV flux.  Since 
spinning dust emission has been observed to be spatially correlated with PAH 
emission (Scaife et al.\ 2010), spinning dust may provide a unique window on 
the small grain population of these disks.  In the context of disk evolution, these
recent measurements conflict with the established 
view that dust grains are expected to grow as disks age.  It may be the case 
that dust fragmentation is important in disks (Dullemond \& Dominik 2005), or 
there exists a separate population of very small carbonaceous grains distinct from 
the MRN distribution (Leger \& Puget 1984; Draine \& Anderson 1985).  This 
second proposition has not only important implications for the study of 
circumstellar disks but also more generally for the complete characterization
of dust and the ISM.  

The arcsecond resolution necessary for these measurements will be achievable 
with several ALMA configurations and Band 1.  From the models of Rafikov 
(2006), the difference between a thermal dust spectrum with $\beta$ $\approx$ 
1 and the predicted spinning dust contribution for a brown dwarf disk would be 
observable at 5 $\sigma$ in a matter of minutes with ALMA Band 1 receivers. 
With longer observation times and consequently higher sensitivity, it will be also
possible to distinguish between different grain size distributions and physical 
conditions within the disk (such as grain electric dipole moments, rotational 
kinematics, optical properties and catalysis of molecule formation).

In summary, spinning dust emission provides a unique insight into the VSG
population under  conditions where it is not possible to observe using mid-IR
emission.  The high resolution and excellent sensitivity of ALMA are ideal for
differentiating the distinct environments where the VSG population resides and
will be crucial for probing star formation and circumstellar regions.  Specifically,
Band 1 receivers will allow routine surveys of the new continuum component
at its spectral maximum.  The smaller minimum baselines of ALMA will make 
it more ideal for probing (especially at southern declinations) the more extended
instances of spinning dust emission, e.g., cores, than the JVLA.  Also, ALMA
Band 1 observations of more compact objects like disks (see \S5.1) will be better
suited for comparison with those at higher frequency bands than those from the
JVLA, given the more similar spatial frequency coverage afforded by observing
from the same latitude.

\subsubsection{Jets from Young Stars}

Radio continuum emission is observed from the jets and winds of young stellar objects and is due to the interaction of free 
electrons, i.e., ``free-free emission." The radio images appear elongated and jet-like and are usually located near the base 
of large optical Herbig-Haro flows (Reipurth \& Bally 2000). These regions usually have only sub-arcsecond sizes, 
indicating the youth of the emitting material and the short dynamical times involved.  The emitted flux is usually weak, with a 
flat to positive spectral index with increasing frequency, and it can be obscured by the stronger thermal emission from dust 
grains at higher frequencies (e.g.,  Anglada 1995).  Multi-wavelength studies of the brightest radio jets at centimeter 
wavelengths trace either earlier and stronger sources or more massive systems. The triple system L1551-IRS 5, one of the 
most studied low-mass systems (Rodriguez et al.\ 1998, 2003; Lim \& Takakuwa 2006), is illustrative of the sub-arcsecond 
scales required (Figure~\ref{fig:shang}).

\begin{figure} 
\centering
\includegraphics[scale=0.8]{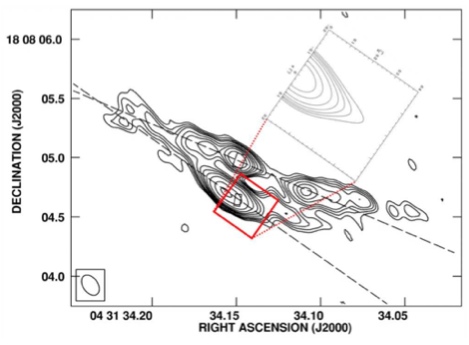}
\caption{{\bf Background}: The VLA+Pie Town continuum image of L1551 IRS 5 at 3.5 cm obtained by Rodriguez et al.\ (2003) in their Figure 1. The size of the beam (0.18 X 0.12\arcsec; P.A. =  35\degree) is shown in the bottom left-hand corner. Black rectangles mark the positions and deconvolved dimensions of the 7 mm compact protoplanetary disks. The dashed lines indicate the position angles of the jet cores. {\bf Inset}: map of the south jet from the X-Wind model convolved with the beam and plotted with the same contour levels from Figure 4 of Shang et al.\ (2004).
}
\label{fig:shang}
\end{figure}

Ground-based, interferometric studies of radio jets provide the best opportunity to resolve the finest scales of the underlying 
source, comparable or better than optical studies of jets by HST.  Such finely detailed images can provide the ability to 
differentiate between theoretical ideas about the nature of these jets, i.e., the launch region, the collimation process, and 
the structure of the inner disks.  Modeling efforts with the radio continuum emission presented in Shang et al.\ (2004) 
demonstrated one such possibility in constraining theoretical parameters using earlier millimeter and centimeter 
interferometers (Figure~\ref{fig:shang}).  Band 1 observations will discriminate between competing jet launch theories
tied to the disk location of the launch point by achieving better than 0.1$^{\prime\prime}$ angular resolution.

The high sensitivity of  ALMA Band 1 observations will also allow detection of radio emission from less luminous sources. 
ALMA will thus have the potential to discover a significant number of new radio jets, providing a catalog from which 
evolutionary changes in the physical properties can be deduced. As well, multi-epoch  surveys will be able to  follow the 
evolution of the freshly ejected material down to a few AU from the driving sources through movies. The 35-52 GHz 
frequency range of Band 1 will show contributions to the observed emission from both the ionized component of the jet and 
the thermal emission from the dust.  These data, together with detailed theoretical modelling will uncover a complete 
understanding of properties of the spectral energy distribution (SED) from the ionized inner regions of young stellar jets.

Relative to the JVLA, Band 1 observations with ALMA may have modest improvements in sensitivity at frequencies in 
common.  Of course, southern sources will be much better observed with ALMA.  Moreover, the wider field-of-view of
ALMA will more easily allow for observations of multiple jets across crowded regions such as within young protoclusters.

\subsubsection{Spatial and Flaring Studies of Sgr~A*}

\begin{figure}
\begin{center}
\includegraphics[scale=0.80]{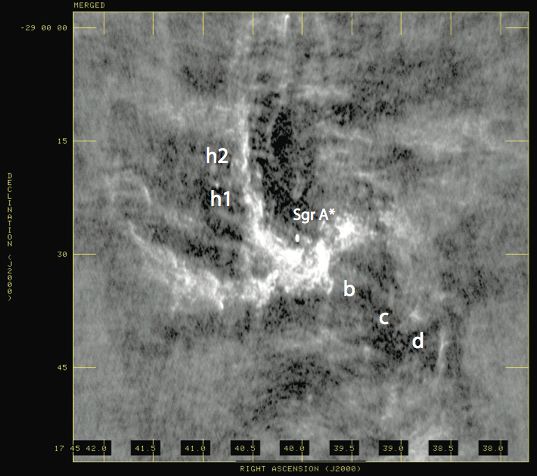}
\caption{ {\it (a) Left} A 22 GHz image of the Sgr~A* region at $0.36''\times0.18''$ resolution (PA=2$^\circ$)  constructed by combining JVLA A- and B- array data.}
\label{fig:sgra22}
\end{center}
\end{figure}

Near-IR and radio observations provide compelling evidence that the compact nonthermal 
radio source Sgr~A* is identified with a 4 $\times$ $ 10^6$ M$_{\odot}$\/ black hole at the
center of the Galaxy (Reid and Brunthaler 2004; Ghez et al. 2008; Gillessen et al. 2009).
It is puzzling, however, that the bolometric luminosity of Sgr~A* due to synchrotron thermal
emission from hot electrons in the magnetized accretion flow is several orders of magnitude
lower than expected from the accretion of stellar winds. There have been two different
approaches to address this puzzling issue.  One is to search for the base of a jet from 
Sgr~A* and identify interaction sites of a jet with the ionized and molecular material
surrounding Sgr~A*.  The other is to study the correlations of the variable emission from
Sgr~A* at centimeter and millimeter bands.  Studies of images and variability are well
suited using ALMA's Band 1 and will be complementary to each other in addressing the
key question as to why Sgr~A* is so underluminous.  Note that Sgr~A* is located at a 
declination of -29$\degree$, making it a more attractive target for ALMA than the JVLA.

Regarding jets, recent JVLA observations at radio wavelengths presented a tantalizing
detection of a jet-like linear feature appearing to emanate from Sgr~A* (Yusef-Zadeh et al.\
2012).  Figure~\ref{fig:sgra22} shows a 23 GHz image of the inner 30$''$ of Sgr~A*. A new
linear feature is noted running diagonally crossing the bright N and W arms of the mini-spiral,
along which several blobs (b, c, d, h1 and h2) are detected. What is interesting about the
direction in which the linear feature is detected is that several radio blobs have X-ray and
FeII/III counterparts also along the axis of the linear structure.  In addition, the extension
of the linear feature appears to be polarized at 8 GHz, suggesting that this feature is a
synchrotron source. The radio-polarized linear jet-like structure is best characterized by
a mildly relativistic jet-driven outflow from Sgr~A*, and an outflow rate $\gamma\dot{M}\sim
10^{-6}$ \msol\, yr$^{-1}$.

The linear arrangements of antennas in the JVLA configurations can lead to linear structures
in the residual beam pattern due to deconvolution errors.  ALMA's configurations, however,
should lead to data with better, more-uniform uv coverage and will establish the reality of the
linear structure.  In particular, Band 1 will be most effective in studying the faint jet-like feature
from Sgr~A*.  Dust emission from the immediate environment of Sgr~A* dominates fluxes at
shorter wavelengths relative to optically thin non-thermal emission from the jet with a steep
energy spectrum.  Thus,  observations with Band 1 are critical for  measuring properly the
morphology, spectral index and polarization characteristics of the jet emanating from Sgr~A*.
Although Sgr~A* is a unique object in the Galaxy, similar motivations also apply to other
non-thermal radio continuum sources such as microquasars, e.g., 1E1740.7-2942, that
have faint radio jets and are located in the inner Galaxy. 

Regarding the correlations of variable emission from Sgr~A*, recent radio measurements
have detected a time delay of $\sim$30 $\pm$ 10 minutes between the peaks of 7 mm and
13 mm radio continuum emission toward Sgr~A* (Yusef-Zadeh et al.\ 2006).  This behaviour
is consistent with a picture of a flare in which the synchrotron emission is initially optically
thick.  Flaring at a given frequency is produced through the adiabatic expansion of an initially
optically thick blob of synchrotron-emitting relativistic electrons.  The intensity grows as the
blob expands, then peaks and declines at each frequency that the blob becomes optically thin. 
This peak first occurs at 43 GHz and then at 22 GHz about 30 minutes later.   Theoretical
light curves of flare emission, as shown in Figure~\ref{fig:synchro}, show that it occurs at high
near-infrared frequencies first and is increasingly delayed at successively lower ALMA
frequencies that are initially optically thick. 

\begin{figure}
\begin{center}
\includegraphics[scale=0.80]{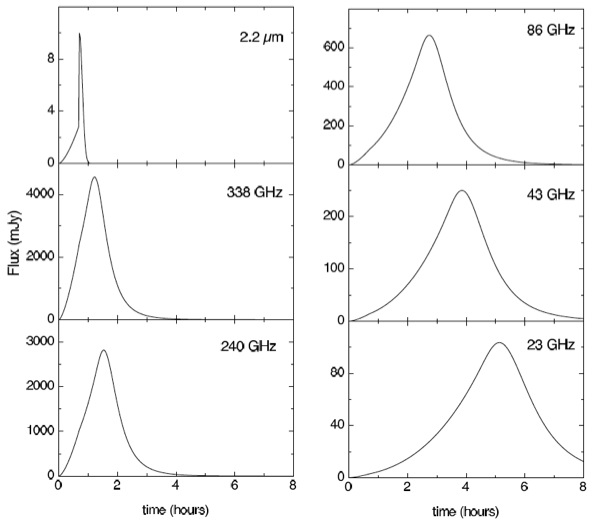}
\caption{ {\it} 
Theoretical light curves of Stokes I for
  optically thick synchrotron emission at five different bands
  corresponding ALMA Bands 3, 6, 7 and 9 as a function of expanding
  blob radius.  These light curves assume an energy power law index
  p=1 where n(E)$\propto$ E$^{-p}$.  
 }
\label{fig:synchro}
\end{center}
\end{figure}

The limited time coverage of JVLA observations at radio wavelengths (due to the low maximum
elevation of Sgr A* at the JVLA) means that there can be a large uncertainty in determining the
underlying background flux level of a particular flare, as well as difficulty identifying flares in
different bands.  Observations of Sgr~A* with a long time coverage using ALMA's Band 1 can fit
the corresponding light curves simultaneously to place much tighter constraints on the derived
physical parameters of the flare emission region. Two parameters of high interest are the
expansion speed of the hot plasma and the initial magnetic field. These quantities characterize
the nature of outflow and cooling processes relevant to millimeter emission.  The fitting of a light
curve at one frequency will automatically generate models for any other frequency.  We should
be able to test the time delay between the peaks of flare emission within Band 1. 

What has emerged from past observing campaigns to study Sgr~A* is that radio, submillimeter,
near-infrared, and X-ray emission can be powerful probes of the evolution of the emitting region
since they are all variable.  We now know that flare emission at infrared wavelengths is due to
optically thin synchrotron emission that is detected when a flare is launched (Eckart et al. 2006).
The relationship between radio and near-infrared/X-ray flare emission has remained unexplored
due the very limited simultaneous time coverage between radio and infrared telescopes.  The
continuous variations of the radio flux on hourly time scales also make the identification of radio
counterparts to infrared flares difficult.  In spite of the limited time coverage, the strong flaring in
near-infrared/X-ray wavelengths has given us an opportunity to examine if there is a correlation
with variability at radio frequencies.  A key motivation for observing Sgr~A* is to compare its flaring
activity with the adiabatic expansion picture. One of the prediction of this model is a time delay
between the peaks of optically thin near-infrared emission and optically thick radio emission, as
discussed above.  From this model,  a near-infrared flare of short duration of 0.5-1 hr is expected
to have a radio counterpart of duration of $\sim2$ hr shifted in time by 3-5 hr.   

\begin{figure}
\begin{center}
\includegraphics[scale=0.75,angle=0]{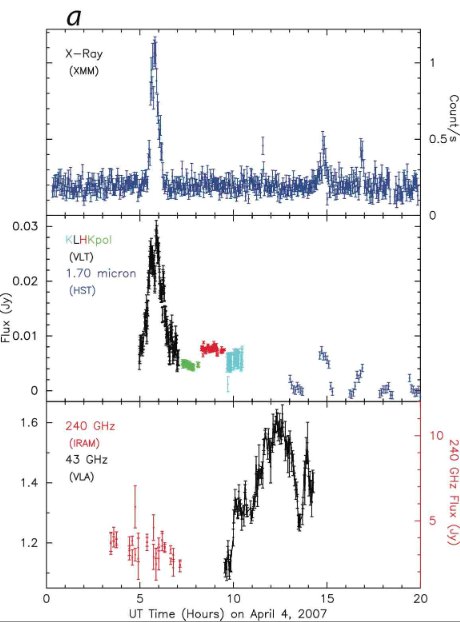}
\caption{The light curves of
Sgr~A* on 2007 April 4 obtained with  XMM in X-rays (top), VLT and HST in NIR
(middle), and IRAM-30m  and  VLA at 240 GHz and 43 GHz, respectively (bottom).
The NIR light curves in the middle panel are represented as H (1.66 $\mu$m) in red,
K$_s$ and K$_s$-polarization mode
(2.12 $\mu$m) in green and light blue, respectively,  L' (3.8$\mu$m) in black
(Dodds-Eden et al. 2009), and
NICMOS of HST in blue at 1.70 $\mu$m. In the bottom panel, red and black colors
represent the 240 GHz and 43 GHz  light curves, respectively.
 }
\label{fig:sgralc}
\end{center}
\end{figure}

Figure~\ref{fig:sgralc} shows composite light curves of Sgr~A* obtained with XMM, VLT, HST,
the IRAM 30-m Telescope, and the VLA on 2007 April 4.  These curves reveal that there was no
significant variation at 240 GHz during the period when the strong near-infrared/X-ray flare took
place.  The IRAM observation shows an average flux of 3.42 Jy $\pm$ 0.26 Jy between 5 hr and
6h UT when the powerful near-infrared flare took place. The millimeter flux is mainly arising from
the quiescent component of Sgr~A*.  Comparing the light curves of the 43 GHz and 240 GHz data,
there is no evidence for a simultaneous radio counterpart to the near-infrared/X-ray flare with no
time delays.  Given the limited coverage in time with the VLA, it is clear that we can not be confident
about the time delay between radio and near-infrared/X-ray peaks.  There is also no overlap in time
between the VLA and Subaru data to test the adiabatic picture of flare emission by making
simultaneous NIR and radio observations.  In future, ALMA and VLT will have the best time overlap
to test this important aspect of flare emission from Sgr~A*.  Although Sgr~A* is a unique object in
the Galaxy, similar arguments could be made for numerous transient sources found in the inner
Galaxy. 

\begin{figure}[t]
\begin{center}
\includegraphics[scale=0.8]{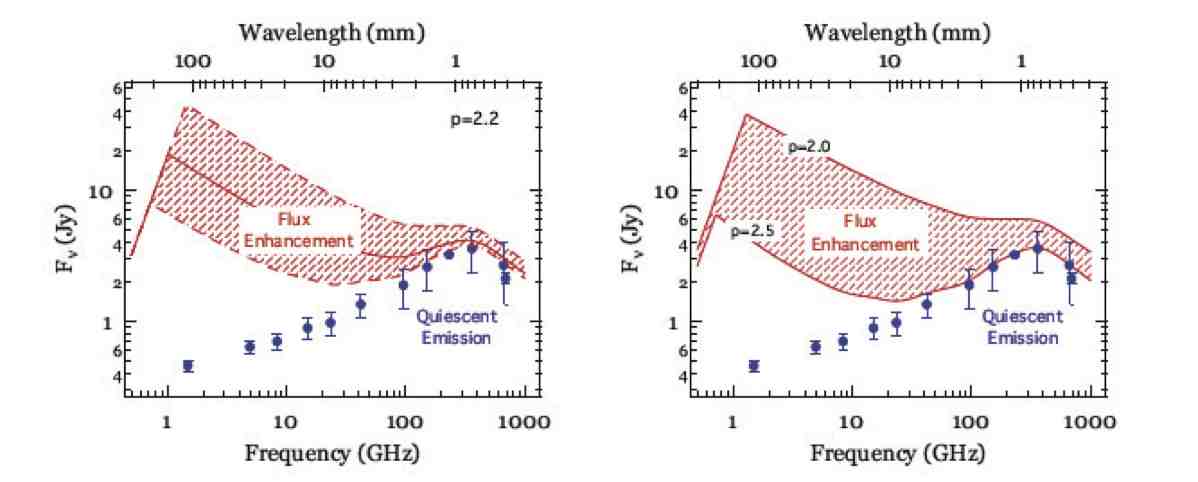}
\caption{
Radio emission as a function of frequency
expected from G2 cloud (red) when compared to
quiescent  emission from Sgr~A*, as shown in blue (Narayan, Ozel, \& Sironi  2012). 
Left and right panels show predictions based on different assumptions
on the energy spectrum of nonthermal particles (p).
}
\label{fig:g2}
\end{center}
\end{figure}

Finally, we note the utility of ALMA Band 1 receivers to trace close encounters of gas clouds with
Sgr~A*.  For example, a 3 M$_{Earth}$ cloud of ionized gas and dust named G2 has been recently
determined to be on a collision course with Sgr~A*.  VLT observations indicate that the G2 cloud
approaches pericenter in mid-2013 and it will be disrupted and portions will likely be accreted by
the massive black hole residing there (Gillessen et al.\ 2012).   At the pericenter distance, the velocity
of the gas cloud will be 5400~km~s$^{-1}$.  Accordingly, the cloud is expected to produce a bow
shock that can easily accelerate electrons into a power-law distribution of index $p = 2.5-3.5$,
assuming standard shock conditions (Narayan et al. 2012).  Depending on $p$, the expected
additional emission from Sgr~A* ranges from 0.6~Jy to 4~Jy, over a dynamical timescale of $\sim$6
months.  The model behind the additional radio emission from the disruption of G2 by the black hole
could have been tested directly with ALMA Band 1 observations.  Though Band 1 receivers will not
be ready for the interaction of G2 with Sgr~A* by 2013, this close encounter is likely not an isolated
event, and future disruptions  of other, similar clouds in the Sgr~A* region by the black hole could be
monitored with Band 1.

In summary, ALMA Band 1 receivers will provide important constraints to models of Sgr~A*, the
supermassive black hole in the center of the Galaxy.  ALMA's southern location will allow for 
improved observations of Sgr~A* than possible at the JVLA site, due to the southern declination
of the object.  For example, the longer time Sgr~A* is present over the horizon improves studies
of variability, and also improves sensitivity and spatial frequency coverage for observations of 
associated phenomena at Band 1 frequencies.

\subsubsection{Acceleration Sites in Solar Flares}

When a solar flare occurs, some of the particles in the corona are accelerated from a few hundred eV up to a few MeV
within less than one second. The non-thermal electrons accelerated by a flare flow along the magnetic field lines of the
flare, emitting microwaves while propagating through the corona.  Finally, they collide with the dense and cool plasma in
the chromosphere and lose the energy by radiation and thermalization.  In most flares, two hard X-Ray (HXR) sources
are observed at the footpoints of the flare loop, and one microwave source is observed around the top of the loop (see
Figure \ref{fig:solar}).  Previous observations of these sources had been done by HXR and microwave solar telescopes
with low spatial resolution (e.g., $\sim10$ arcsec) and low dynamic range (10--100).  Hence, it has been hard to investigate
the structures and time evolution of the sources behind particle acceleration, especially since we do not yet know where
the acceleration site is in a flare.  Some indirect evidence suggests that the acceleration site is located above the flare loop,
in a location filled with $\sim10$ MK thermal plasma (Masuda et al. 1994, Aschwanden et al. 1996, Sui and Holman 2003),
but there is no direct evidence yet.  Currently, it is also impossible to investigate the the relationship of the acceleration 
site with the thermal structures, like the in-flow of magnetic reconnection detected by the EUV observations (Yokoyama
et al.\ 2001). Therefore, there has been no significant progress in the study of the particle acceleration in the last decade. 

\begin{figure}
\begin{center}
\includegraphics[scale=0.6]{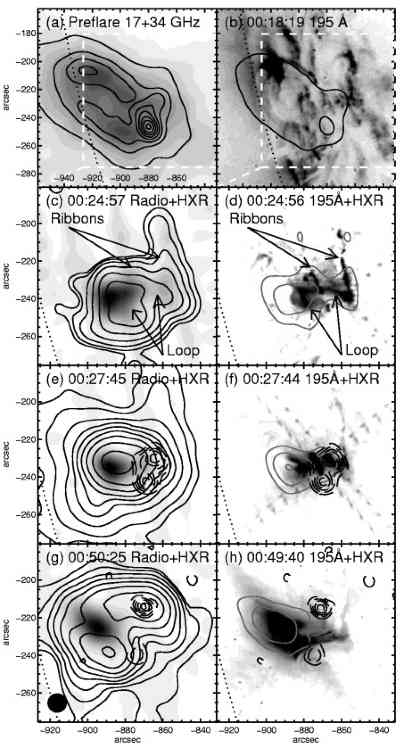}
\caption{Images of a solar flare at X-ray, EUV, and radio wavelengths. The top
  row panels show radio and EUV images in the pre-flare phase.  On the left are
  17 GHz contours overlaid on a greyscale 34 GHz image (both averaged over
  the period 23:00-00:15 UT), while the right panel shows a 195 \AA\ image from
  00:18:19 UT together with two 17 GHz contours for context.  The remaining
  rows of panels show the 96$^{\prime\prime}$ $\times$ 96$^{\prime\prime}$ region
  outlined in the pre-flare images.  The left panels show the RHESSI greyscale
  image of 12-20 keV HXR overlaid with 17 GHz total intensity radio contours (solid curves)
  and RHESSI 100-150 keV HXR contours (dashed curves).   The right panels
  show a 195 \AA\ image of the same region overlaid with solid grey contours for
  the RHESSI 12-20 keV HXR and dashed black contours for the RHESSI 100-150
  keV HXR.  The panel labels refer to the times of the 17 GHz images (left) and
  the TRACE images (right). 
  Figure from White et al.\ (2003).}
\label{fig:solar}
\end{center}
\end{figure}

Breakthroughs in the study of the particle acceleration in a solar flare may be possible by solar ALMA observations even
with ALMA's current specs, because its spatial resolutions and dynamic ranges are one order magnitude higher than the
current solar HXR and microwave telescopes.  Nevertheless, the possibility is very tiny for two important reasons: \textit{1)}
the field-of-view of ALMA Band 3, the presently lowest observing frequency receiver of ALMA, is about 60$^{\prime\prime}$.
That field-of-view is not large enough for most flare observations and also it would be very hard to observe simultaneously
the region above the flare loop predicted to be the acceleration site and the flare loop itself.  Moreover, the size of the
field-of-view is directly related to the possibility of observing flares, since the duration of solar observations by ALMA is
limited. \textit{2)} If the acceleration site is above the flare loop, as suggested by indirect evidence, we can easily infer
that the magnetic field strengths at the site is a few tens of Gauss.  The emissivity of the microwaves emitted by the
gyro-synchrotoron mechanism, however, strongly depends on the magnetic field strength. Therefore, emission at frequencies
of 230 GHz and higher from the acceleration site is very weak.   Such high frequency emission has been detected only
from the main sources of large flares by submillimeter single-dish observations (e.g., Kaufmann, et al. 2004).  Therefore,
a lower frequency band with the high spatial resolution and dynamic range of ALMA is needed to observe the non-thermal
emission from the acceleration site.  Flare observations with ALMA Band 1, with a single-pointing field-of-view of about
100$^{\prime\prime}$ in the 35--50 GHz frequency range, will obtain significantly better results for the particle acceleration
studies of a solar flare. If the Band 1 receiver has also the capability to observe circular polarization, even higher scientific
returns will be achieved, because the circular polarization of the gyro-synchrotoron emission will reveal the magnetic field
strength of the emitting region.

The JVLA can also observe the Sun at similar frequencies as those of Band 1, but JVLA solar observations have several
disadvantages.  First, the JVLA has a more reduced $u$--$v$ coverage.  To synthesize a solar image, snapshot data are
needed because the non-thermal emission from a solar flare changes within less than one second.  Hence, ALMA's larger
number of baselines means that a larger number of data points will be instantaneously sampled on the u-v plane.  Second,
since the JVLA antennas are larger than the ALMA antennas and the JVLA cannot sample as many short spacings, the
maximum angular scale observable with the JVLA is $\sim32^{\prime\prime}$, making it harder to reconstruct flare loops
than with ALMA.  Finally, the field-of-view of the JVLA, $\sim60^{\prime\prime}$, is relatively small.

The total flux of gyro-syncrhotron emission emitted from a solar flare follows a power-law distribution with frequency in
the optically-thin frequency range, so lower frequency observations are more sensitive in detecting flares.  The typical
turnover frequency of flares is about 10 GHz.  Therefore, the total flux of emission in the Band 1 frequency range is one
to two orders of magnitude larger than that  in Band 3.  Nobeyama polarimeter data have shown that the total flux average
from 700 solar flares at 35 GHz is 46.3 SFU ($4.63\times10^5$ Jy).  Special care has to be taken to deal with such a large
input flux.

\subsubsection{Pulsar Wind Nebulae}  

Pulsars generate magnetized particle winds that inflate an expanding 
bubble called a pulsar wind nebula (PWN) whose outer edge is confined by 
the slowly expanding supernova ejecta. Electrons and positrons are 
accelerated at the termination shock some $0.1\,$pc distant from the 
pulsar. Those relativistic particles interact with the magnetic field inside the 
wind-blown bubble to produce synchrotron emission across the entire 
electromagnetic spectrum. Particles accelerated at the shock form 
toroidal structures, known as wisps, and some of them are collimated 
along the rotation axis of the pulsar, contributing to the formation of 
jet-like features. The synchrotron emission structure in the post-shock 
and jet regions provide direct insight on the particle acceleration 
process, magnetic collimation, and the magnetization properties of the 
winds in PWNe. These observations have so far (except for the Crab 
Nebula) been limited to X-ray wavelengths with the {\it Chandra}\/
satellite (e.g., Helfand et al.\ 2001).

\begin{figure}
\centering
\includegraphics[scale=0.8]{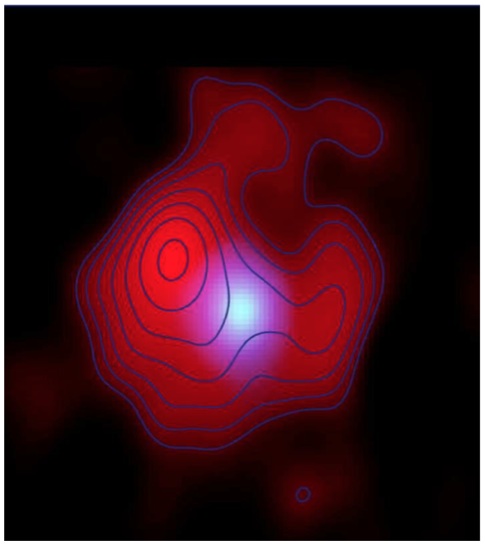}
\caption{
Two-colour VLBI image of SN 1986J highlighting the emergence of a central
component. The red colour and the contours represent the 5.0 GHz radio
brightness. The contours are drawn at 11.3, 16. 22.6É90.5\% of the peak
brightness of 0.55 mJy/bm. The blue to white colours show the 15 GHz brightness
of the compact, central component. The scale is given by the width of the
picture of 9 mas. North is up and east to the left. For more information on the emergence 
of the compact source, see Bietenholz et al.\ (2004). }
\label{fig:cont2}
\end{figure}

ALMA has the sensitivity and resolution necessary to detect PWNe features 
at high radio frequencies, where we can detect the emission from 
relativistic particles that have much longer lifetimes than in
X-rays.
At cm/mm-wavelengths, flat-spectrum synchrotron PWNe
stand out over steep-spectrum SNRs 
(e.g., as seen in the Vela PWN (Hales et al.\ 2004), discussed
in \S~6.1.6 below, and illustrated in Figure~\ref{fig:cont2} (Bietenholz
et al.\ 2004)) with minimal
confusion from the Rayleigh-Jeans tail of submm dust. ALMA Band 1
receivers
will allow observations in the frequency regime where PWNe dominate, and
bridge an important gap in frequency coverage, where spectral features
such as power-law breaks occur and linear polarization observations
do not suffer from significant Faraday rotation.  Here, even the modest
improvements in sensitivity of ALMA in Band 1 over the JVLA at similar
frequencies will be important.  Also, of course, southern PWNe will be
much better probed with ALMA.

\subsubsection{Radio Supernovae} 

Radio supernovae occur when the blast wave of a core-collapse supernova (SN)
sweeps through the slowly expanding wind left over from the progenitor red
supergiant. Particle acceleration and magnetic field amplification lead to
synchrotron radiation in a shell bounded by the forward and reverse shocks
(Chevalier 1982).  In general, free-free absorption of the radiation in the
ionized foreground medium coupled with the expansion of the SN causes the radio
light curve first to rise at high frequencies and subsequently at progressively
lower frequencies while the optical depth decreases. When the optical depth has
reached approximately unity, the radio light curve peaks and decreases
thereafter (e.g., Weiler et al.\ 2002). These characteristics allow estimates to
be made of the density profiles of the expanding ejecta and the circumstellar
medium and also of the mass loss of the progenitor. Resolved images of SNe
provide information, e.g., on the structure of the shell, size, expansion
velocity, age, deceleration, and magnetic field, in addition to refined
estimates of the density profiles and the mass loss (Bartel et al.\ 2002).
Radio observations of SNe can be regarded as a time machine, where the history of
the mass loss of the progenitor is recorded tens of thousands of years before
the star died.
Finally, the SN images can be used to make a movie of the expanding shell of
radio emission and to obtain a geometric estimate of the distance to the host
galaxy (Bartel et al.\ 2007).

ALMA Band 1 receivers will allow exciting science to be done in the areas of radio
light curve measurements, imaging of a nearby SN and, in conjunction with VLBI,
imaging of more distant SNe. Depending on the medium, the delay between the
peak of the radio light curve at 20 cm and 1 cm can be as long as 10 years, as
for instance was the case of SN 1996cr (Bauer et al.\ 2008). Absorption
can also occur in the source itself. In case of SN 1986J, a new component
appeared in the radio spectrum and in the VLBI images about 20 years after the
explosion and then only at or around 1 cm wavelength. The component is located
in the projected center of the shell-like structure of the SN and may be
emission from a very dense clump fortuitously close to that center, or possibly
from a pulsar wind nebula in the physical center of the shell (Figure~\ref{fig:cont2},
Bietenholz et al.\  2004, 2010).  Observations in Band 1 minimize the absorption effect
relative to observations at longer wavelengths and thus allow investigations of
SNe at the earliest times without compromising too much on the signal to noise
ratio of a source with a steep spectrum. ALMA with Band 1 receivers has the
sensitivity to measure the radio light curves of 10s to 100 SNe. In addition, ALMA
may be then also particularly sensitive in finding ``SN factories" in starburst galaxies
(e.g., Lonsdale et al.\ 2006) where relatively large opacities would otherwise
hinder or prevent discovery.

ALMA with Band 1 receivers will allow high-dynamic range images of SN 1987A in
the Large Magellanic Cloud with a resolution of about 300 FWHM beams across the
area of the shell in 2014.   Such data would be a significant improvement over presently
obtainable images (Gaensler et al.\ 2007; Laki{\'c}evi{\'c} et al.\ 2012).  Also, since the
size of the SN increases by one Band 1 FWHM beam width per 3 years, the expansion
of the shell can be monitored accurately and in detail, making this SN an important
target for ALMA.  

In summary, ALMA Band 1 receivers could make strides in observing high-frequency
synchrotron from supernovae, allowing important measurements of their properties.
ALMA's location in the southern hemisphere makes investigations of southern SNe 
(expecially SN 1987A) especially compelling.  Note that ALMA's southern berth also
would make it an important element of VLBI arrays operating in Band 1, providing 
southern baselines and high sensitivity.  Previous SN VLBI observations at 1 cm
wavelength have provided clues about physical conditions at the earliest times
after the transition from opaqueness to transparency, and SN VLBI with Band 1
will surely focus on this area of research.

\subsubsection{X-ray Binaries}

X-ray binaries (i.e., binary star systems with either a neutron star or a black 
hole accreting from a close companion) frequently show jet emission.  Most of 
these systems are transients. Typically, 1-2 black hole X-ray binaries undergo 
a transient outburst per year, while neutron stars outburst at a slightly higher 
rate.  Outbursts typically last several months (although there are some which
are both considerably longer or shorter), and during outbursts, X-ray luminosities
can change by as much as 7 orders of magnitude. The radio luminosities of
systems seen to date correlate well with the hard X-ray luminosities (i.e., those
above $\sim$20 keV), albeit with considerable, yet poorly understood scatter.

When the X-ray spectra become dominated by thermal X-ray emission, the radio 
emission often turns off (e.g., Tananbaum et al.\ 1972; Fender et al.\ 1999), 
but the extent to which the flux turns down is still poorly constrained. This 
turndown is not seen in neutron star X-ray binaries (Migliari et al.\ 2004). 
The reduced radio emission in black hole X-ray binaries when they have soft 
X-ray spectra can be explained by models of jet production in which the jet 
power scales with the polodial component of the magnetic field of the accretion 
flow (e.g., Livio, Ogilvie \& Pringle 1999), and may have implications for 
the radio loud/quiet quasar dichotomy (e.g., Meier 1999; Maccarone, 
Gallo \& Fender 2003).  The still-present radio emission from neutron stars 
in their soft state may be indicating that the neutron star boundary layers 
play an important role in powering jets (Maccarone 2008).  The soft states
of X-ray transients are short-lived.  During them, there may be decaying 
emission from transient radio flares launched during the state transitions. 
Therefore, to place better upper limits on the radio jets produced during 
the soft state, a high sensitivity, high frequency system with a very high duty
cycle is needed.

The radio properties of X-ray binaries with neutron star primaries are much 
more poorly understood than those of black hole X-ray binaries.  This situation
is partially because the neutron star X-ray binaries are fainter in X-rays than 
are the black hole X-ray binaries. There is, however, additionally some 
evidence that neutron star X-ray binaries show a steeper relation between X-ray 
luminosity and radio luminosity than do the black hole X-ray binaries, with 
$L_R \propto L_X^{0.7}$ for the black holes and $L_R \propto L_X^{1.4}$ for 
the neutron stars.  This difference may be explained if the neutron stars are 
radiatively efficient (i.e., with the X-ray luminosity scaling with the accretion
rate) while the black holes are not (i.e., with the X-ray luminosity scaling with the
square of the accretion rate, as has been proposed by Narayan \& Yi 1994) -- 
see Koerding et al.\ (2006). Radio/X-ray correlations for neutron star X-ray
binaries are, to date, based on small numbers of data points from few sources,
and the most recent work (Tudose et al.\ 2009) indicates that the situation may
be far more complex than the picture presented above. 

In summary, Band 1 frequencies are important for resolving the relationship
between radio and X-ray flares in transient events from neutron star and black
hole binaries.  ALMA with Band 1 receivers would provide the ability to catch
such events at southern declinations.  ALMA's high sensitivity is especially
important to constrain the downturns at radio wavelengths seen in many events.

\subsection{Line Observations with ALMA Band 1}

As with the continuum science cases, numerous examples of scientific opportunity will be 
available to ALMA users interested in the numerous lines located in the Band 1 frequency
range from molecular rotational transitions and radio recombination lines.  Here we discuss
some science cases that involve high sensitivity observations of lines, including studies of
(1) chemical differentiation in cloud cores; (2) the chemistry of complex carbon-chain
molecules; (3) ionized gas in the dusty nuclei of starburst galaxies; (4) the photoevaporation
of protoplanetary disks; (5) inflows and outflows from HII regions; (6) masers; (7) magnetic
field strengths in dense gas; (8)  molecular outflows from young stars; (9) the
co-evolution of star formation and active galactic nuclei; and (10) the molecular gas
content of star-forming galaxies at $z$ $\sim$ 2.

\subsubsection{Fine Structure of Chemical Differentiation in Cloud Cores}

Previous single-dish millimeter molecular line observations have found that molecular 
abundance distributions differ significantly between individual dark cloud cores.  A widely
accepted interpretation of this chemical differentiation is that there exists non-equilibrium 
gas-phase chemical evolution through ion-molecule reactions within dark cloud cores.
Younger cores are rich  in ``early-type" carbon-chain molecules such as CCS and HC$_3$N, 
while more evolved cores, closer to protostellar formation via gravitational collapse, are 
rich in ``late-type" molecules such as NH$_3$ and SO (Suzuki et al.~1992). 
Recent high-resolution millimeter-line observations, however, have revealed that
there are even finer variations of molecular distributions within cores down to $\sim$3000 AU
scales, and that these fine-scale chemical fluctuations cannot be explained by the simple
scenario of chemical evolution of cores (Takakuwa et al.\ 2003, Buckle et al.\ 2006).  The
explanation suggested for this behaviour is that there is first molecular depletion onto grain
surfaces in these regions and then subsequent reaction and desorption of molecules back
to the gas phase through clump-to-clump collisions or energy injection from newly formed
protostars (e.g., Buckle et al.\ 2006).  The molecules that can differentiate between regions
with ``early--type" chemistry, before any collapse of a protostellar object, and the ``late--type"
chemistry, apparent after the formation of a protostellar core, have their ground-state (strongest)
transitions in ALMA Band 1. These heavy saturated organic molecules can only be formed
on the surfaces of dust grains, and so their appearance in the interstellar medium signals the
presence of a central heating source, likely a protostar.  ALMA Band 1 receivers will provide
the most sensitive test of when a central heating source turns on, since ALMA will then have
the resolution and sensitivity to detect the presence of these complex molecules within a
dense core of more diffuse, unprocessed gas.

Other recent work (see Garrod, Weaver \& Herbst 2008 and references therein)
has shown some surprising detections of saturated complex organic molecules
around apparently quiescent dust cores, consistent with model predictions for the
``warm-up" chemistry expected when a core is undergoing gravitational collapse
and forming an internal heating source.  According to models, a later stage in this
sequence occurs when complex saturated molecules produced on grain surfaces
react as the gas warms up, producing ``hot core" chemistry, with even more
complex products.

In summary, ALMA Band 1 receivers will allow probes of the smallest length scales
of chemical variation in cloud cores to clarify the relationship among different
molecular abundance distributions (in conjunction with chemical models).  These 
projects will require both ALMA's excellent spatial resolution and in particular its ability
to recover the larger-scale structure of cores through observations with the ACA.
Indeed, ALMA's higher sensitivity to extended, surface brightness emission and high
fidelity make observations of such lines preferable to observations of them with the
JVLA.  Also, ALMA Band 1 will likely include 50-52 GHz, a frequency range unavailable
with the JVLA that contains many interesting lines, including C$_{3}$H$_{2}$
1$_{1,1}$--0$_{0,0}$ at 51.8 GHz.  Table~\ref{table:mol} lists some molecular
transitions needed for the chemical studies within these clouds that are observable
over 35-52 GHz.  

\begin{table*}[h]

\caption{Molecular Transitions between 35 GHz and 52 GHz}
\label{table:mol}

\begin{center}
\begin{tabular}{lcr}
\hline
SO & 2$_3$--2$_2$& 36.202040 GHz\\
HC$_3$N & 4--3& 36.392332 GHz\\
HCS$^+$ &1--0 & 42.674205 GHz\\
SiO & 1--0 & 43.42376 GHz\\
HC$_5$N & 17--16 & 45.264721 GHz\\
CCS & 4$_3$--3$_2$& 45.379033 GHz\\ 
HC$_3$N & 5--4 & 45.490316 GHz\\
CCCS & 8--7& 46.245621 GHz\\
C$_3$H$_2$ & 2$_{1,1}$--2$_{0,2}$& 46.755621 GHz\\
C$^{34}$S & 1--0& 48.206956 GHz\\
CH$_3$OH & 1$_0$--0$_0$& 48.372467 GHz\\
CS & 1--0 & 48.99096 GHz\\
HDO & 3$_{2,1}$--3$_{2,2}$ & 50.23630 GHz\\
HC$_5$N & 19--18 & 50.58982 GHz\\
DC$_{3}$N & 6--5 & 50.65860 GHz\\
O$_{2}$ & N=35-35, J=35-34 & 50.98773 GHz\\
CH$_{3}$CHO & 1(1,1)-0(0,0) & 51.37391 GHz\\
NH$_{2}$D & 1(1,0)--1(1,1) & 51.47845 GHz\\
CH$_{2}$CHCHO & 1$_{11}$--0$_{00}$ & 51.59607 GHz\\
C$_3$H$_2$ & 1$_{1,1}$--0$_{0,0}$& 51.841418 GHz\\
\hline
\end{tabular}
\end{center}
\end{table*}

\subsubsection{Complex Carbon Chain Molecules}

Band 1 receivers will provide the opportunity to search with ALMA for new
complex organic molecules, including the amino acids and sugars from which
life on Earth may have originally evolved.  In addition, these complex molecules
provide a powerful tool for understanding star formation and the processes
surrounding it.

There are several reasons why Band 1 is the best place to search for
complex molecules.  First, the heavier a molecule, the lower will be its 
rotational transition frequencies. The many abundant lighter molecules
(e.g., CO, HCN, CN) have their lowest transitions in Band 3, and so do
not appear at all in Band 1. Therefore, Band 1 does not suffer from
contamination from these common molecules, and so line confusion
is much less of a problem.  Second, system temperatures at Band 1
frequencies will be significantly lower than in higher bands, giving extra
sensitivity to detect weak transitions from less abundant complex molecules,
such as glycolaldehyde, the simple sugar known to exist in the interstellar
medium.  Table~\ref{table:ccm} lists some complex carbon-chain
molecules whose transitions have been already detected in the ISM.
Note that searches for complex molecules can be made with Band 1
also using lines in absorption against bright background objects like,
e.g., young stars or quasars.

There is now a significant body of evidence to suggest that complex
biological molecules, such as amino acids and sugars needed for
evolution of life on Earth, evolved in the interstellar medium (e.g.,
see Holtom et al.\ 2005; Hunt-Cunningham \& Jones 2004; Bailey 
et al.\ 1998).  Band 1 receivers will be one of the best instruments in
the world to test this hypothesis observationally. 

As with the molecular transitions described in \S 6.2.1, ALMA's 
sensitivity to low surface brightness line emission through the smaller
minimum baselines of the 12-m Array and the ACA itself makes
exploring complex carbon-chain molecular chemistry preferable
with ALMA than the JVLA over 35-50 GHz.  In addition, the likely
addition of 50-52 GHz to the Band 1 frequency range is not available
at the JVLA.

\begin{table*}[h]

\caption{Some detected ISM complex carbon chain molecules}
\label{table:ccm}

\begin{center}

\begin{tabular}{lr}

\hline

CH$_2$CHCN & propenitrile\\

CH$_2$CNH & ketenimine\\

CH$_3$C$_4$H & methyldiacetylene\\

CH$_3$CCCN & methyl cyanoacetylene\\

CH$_3$CH$_2$CN & ethyl cyanide\\

CH$_3$CHO & acetaldehyde\\

CH$_3$CONH$_2$ & acetamide\\

CH$_3$OCH$_3$ & ethyl butyl ether\\

CH$_3$OCHO & methyl formate\\

C$_6$H$^-$ & hexatriyne anion\\

C$_8$H & octatetraynyl\\

H$_2$CCCC & cumulene carbene\\

HCCCNH$^+$ & \nodata \\

\hline
\end{tabular}
\end{center}
\end{table*}

\subsubsection{Radio Recombination Lines}

In the radio and submillimeter, we have access to an extinction-free
ionized gas tracer: radio recombination lines (RRLs). These lines can
measure the density, filling factor, temperature, and kinematics of
the ionized gas in young star-forming regions that are still heavily
obscured by dust. Measuring the properties of the ionized gas in these
regions allows us to probe the properties of the interstellar medium
and the stars in a very early stage of star formation.  RRLs in the 
ALMA Band 1 frequency range (e.g., H53$\alpha$ at 43.309 GHz)
trace ionized gas with densities of $10^4 \ {\rm cm}^{-3}$, which is
similiar to the densities of young HII regions (Churchwell 2002).

Using RRLs detected in ALMA Band 1, we can:
\begin{itemize}
\item measure the properties of the ionized gas and young massive
  stars in the dusty nuclei of starburst galaxies (see Figure
  \ref{fig:evla_results}; Kepley et al.\ 2011),
 \item detect the photoevaporation of protoplanetary disks
  (Pascucci, Gorti \& Hollenbach 2012), and
\item quantify the properties of inflows and outflows from HII regions
  (Peters et al.\ 2012) and possibly gas ionized by jets from
  young stars (Shepherd et al.\ 2013).
\end{itemize}

In the past, RRLs were difficult to observe -- particularly in
external galaxies -- because they are faint and broad lines. Today,
the high sensitivity and wide bandwidths of facilities like ALMA make
RRLs more accessible. The wide band widths also allow us to stack
RRLs. RRL properties change slowly with frequency, so stacking all
RRLs observed within a band improves the sensitivity of the
observations without increasing the observing time or affecting the
properties of the line.

RRLs are brighter at higher frequencies, but they also are further
apart in frequency space. ALMA Band 1 frequencies are ideal for RRL
detection because the lines are bright and we can detect 3-4 lines in
the 8 GHz of bandwidth provided by the ALMA correlator. At lower
frequencies, the lines will be fainter; at higher frequencies, we
cannot stack as many lines.

\begin{figure}
\centering
\includegraphics[width=1.0\textwidth]{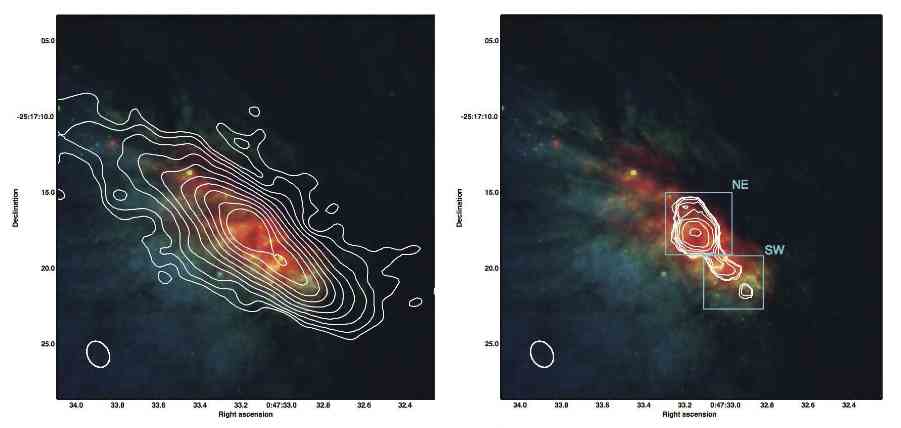}
\caption{RRLs can measure the ionized gas properties in the dusty
  nuclei of starburst galaxies. The left panel shows JVLA observations
  of the 1cm continuum emission, which is mostly free-free emission,
  from the nuclear starburst of the edge-on galaxy NGC 253. The right
  panel shows JVLA observations of the \hrrl{58} emission from the
  same galaxy. The background image shows optical {\em HST}
  images. Paschen $\alpha$ is red, I band is green, and B band is
  blue. Figure from Kepley et al. (2011).}
\label{fig:evla_results}
\end{figure}

Modeling RRL emission requires a sensitive measurement of the
free-free continuum. At the ALMA Band 1 frequencies, the free-free
continuum begins to dominate over the synchrotron and dust continua,
making measuring the free-free component straightforward. Modeling
RRLs at frequencies higher than $\sim$100~GHz requires disentangling
free-free and dust emission.

In summary, ALMA Band 1 receivers will allow the RRLs in its frequency
range to be observed towards many possible targets, including the dusty
nuclei of starburst galaxies, photoevaporating disks, and HII regions.  The
southern location of ALMA will allow southern examples of these sources
to be easily observed to high sensitivity.

\subsubsection{Maser Science}

Masers (Microwave Amplifications by Stimulated Emission of Radiation) frequently occur in regions of active star formation, from molecular transitions whose populations are either radiatively or collisionally inverted. A photon emitted from this material will interact with other excited molecules along its path, stimulating further emission of identical photons. This process leads to the creation of a highly directional beam that has sufficient intensity to be detected at very large distances.

Masers are observed from a variety of molecular and atomic species and each serves as a signpost for a specific phenomenon, a property which renders masers powerful astrophysical tools (Menten 2007). More precisely, masers are formed under specific conditions, and the detection of maser emission therefore suggests that physical conditions (e.g., temperature, density, and molecular abundance) in the region where the maser forms lie within a defined range (c.f., Cohen 1995, Ellingsen 2004, and references therein). Therefore, interferometric blind and targeted surveys of maser species can lead to the detection of objects at interesting evolutionary phases (Ellingsen 2007).

\begin{table*}[h]
\caption{ALMA bands with known maser lines (Menten 2007)}
\label{table:maser1}
\newcommand{\m}{\hphantom{$-$}}
\newcommand{\cc}[1]{\multicolumn{1}{c}{#1}}
\renewcommand{\tabcolsep}{2.0pc} 
\renewcommand{\arraystretch}{1.2} 
\begin{center}
\begin{tabular}{@{}lclclcl}
\hline
 Species  &  ALMA Bands\\
\hline
 \m H$_2$O & \m B3, B5, B6, B7, B8, B9\\
 \m CH$_3$OH & \m B1, B3, B4, B6\\
 \m SiO& \m B1, B2, B3, B4, B5, B6, B7\\
 \m HCN & \m B3, B4, B6, B7, B9\\
 \hline
\end{tabular}\\[0.1pt]\
\end{center}
\end{table*}

Theoretical models of masers strongly depend on physical conditions as well as the geometry of the maser source. A 
successful model should be able to reproduce observational characteristics of observed maser lines but also to predict new 
maser transitions (e.g., the models of Sobolev 1997 for Class II methanol masers and Neufeld 1991 for water masers). In 
that respect, interferometry is essential for the successful search of candidate lines and confirmation of their maser nature.  
ALMA, in particular, will resolve closely spaced maser spots and help further establish precise models of masing sources 
by determining if the detected maser signals are associated with thermal emission (Sobolev 1999), which is essential
for improving theoretical models. With Band 1, ALMA will cover an even wider frequency range, making it ideal for
multi-transition observations of various maser species across the millimeter and  submillimeter windows.  Examples of 
species with observed maser radiation in the different ALMA bands are given in Table \ref{table:maser1}, while Tables \ref
{table:maser2} \& \ref{table:maser3} list SiO and methanol maser transitions that have been observed or predicted to be 
within Band 1.  

Maser radiation can be linearly or circularly polarized depending on the magnetic properties of the molecule. Polarimetric 
studies of maser radiation with interferometers can therefore yield information on the morphology of the magnetic field 
threading the region on small scales, with the plane-of-sky and line-of-sight components of the field being probed using 
linear and circular polarization measurements, respectively (e.g., see Harvey-Smith 2008, Vlemmings 2006). Polarization 
data are essential for improving on the theory of maser polarization first introduced by Goldreich (1973a), which applies to a 
linear maser region, a constant magnetic field, the simplest energy states for a masing transition, and asymptotic limits. 
Observations at higher spatial resolution are needed to verify and improve on more realistic and extensive models (Watson 
2008).

In summary, the ALMA Band 1 frequency range contains numerous CH$_{3}$OH and SiO maser lines that can be 
observed to trace very distinct conditions in the ISM and probe maser production mechanisms.  With ALMA's high
resolutions and sensitivities in the south, the Band 1 receivers will be able to trace easily masers from southern
sources, and provide highly complementary data to masers observed in the higher frequency ALMA Bands.

\begin{table*}[h]
\caption{Observed SiO maser lines in the Band 1 of ALMA (Menten 2007).}
\label{table:maser2}
\newcommand{\m}{\hphantom{$-$}}
\newcommand{\cc}[1]{\multicolumn{1}{c}{#1}}
\renewcommand{\tabcolsep}{2.0pc} 
\renewcommand{\arraystretch}{1.2} 
\begin{center}
\begin{tabular}{@{}lclclcl}
\hline
 Transitions  &  Frequency (GHz)\\
\hline
\m v=0 (J= 1 $\to$ 0) & \m 42.373359\\
  \m v=3 (J= 1 $\to$ 0) & \m 42.519373\\
\m v=2 (J= 1 $\to$ 0) & \m 42.820582\\
\m v=0 (J= 1 $\to$ 0) & \m 42.879916\\
 \m  v=1 (J= 1 $\to$ 0) & \m 43.122079 \\
\m v=0 (J= 1 $\to$ 0) & \m 43.423585\\
 \hline
\end{tabular}\\[0.1pt]\
\end{center}
\end{table*}

\begin{table*}[htb]
\caption{Observed (Menten 2007) and predicted (designated with a star, Cragg et al. 2005) methanol maser lines in Band 1}
\label{table:maser3}
\newcommand{\m}{\hphantom{$-$}}
\newcommand{\cc}[1]{\multicolumn{1}{c}{#1}}
\renewcommand{\tabcolsep}{2.0pc} 
\renewcommand{\arraystretch}{1.2} 
\begin{center}
\begin{tabular}{@{}lclclcl}
\hline
 Transitions  &  Frequency (GHz)\\
\hline
 \m  4(-1) $\to$ 3(0)E & \m 36.1693 \\
 \m 7(-2) $\to$ 8(-1)E & \m 37.7037\\
 \m 6(2) $\to$ 5(3)A$^+$ & \m 38.2933\\
\m 6(2) $\to$ 5(3)A$^-$ & \m 38.4527\\
\m 7(0) $\to$ 6(1)A$^+$ & \m 44.0694\\
\m 2(0) $\to$ 3(1)E  $^*$ & \m 44.9558\\
\m 9(3) $\to$ 10(2)E $^{*}$ & \m 45.8436\\
\hline
\end{tabular}\\[0.1pt]\
\end{center}
\end{table*}

\newpage

\subsubsection{Magnetic Field Strengths from Zeeman Measurements}

Magnetic fields are believed to play a crucial role in the star
formation process.  Various theoretical and numerical studies explain
how magnetic fields can account for the support of clouds against
self-gravity, the formation of cloud cores, the persistence of
supersonic line widths, and the low specific angular momentum of cloud
cores and stars (McKee \& Ostriker 2007).  The Òstandard
modelÓ  suggests that the initial mass-to-(magnetic) flux ratio,
M/$\Phi_{init}$, is the key parameter governing the fate of molecular
cores.  Namely, if the M/$\Phi_{init}$ of a core is greater than the critical
value, the core will collapse and form stars on short time scales, but
for cores with M/$\Phi_{init}$ smaller than the critical value the
process of ambipolar diffusion will take a long time to reduce the magnetic
pressure (Mouschovias \& Spitzer 1976; Shu et al.\ 1987).  On the
other hand, recent MHD simulations suggest that turbulence can control
the formation of clouds and cores.  In such cases, the mass-to-flux
ratio in the center of a collapsing core will be larger than that in
its envelope, the opposite of the ambipolar diffusion results (Dib et
al.\ 2007).  Therefore, measuring the magnetic field strengths and the
mass-to-flux ratios in the core and envelope provide a critical test
for star formation theories.

Despite its central importance, the magnetic field is the most poorly
measured parameter in the star formation process.  The main problem is
that magnetic fields can be measured only via polarized radiation,
which requires extremely high sensitivity for detections.  As a
result, the observed data on magnetic fields is 
sparse compared with those related to the densities,
temperatures, and kinematics in star-forming cores.  The large
collecting area of ALMA provides the best opportunity to
resolve the sensitivity problem for magnetic field measurements.

The key to determining mass-to-flux ratios is 
the measurement of the strength of magnetic fields.
This measurement can be made {\it directly} through detection
of the Zeeman effect in spectral lines.
Observations of Zeeman splitting involve detecting the small
difference between left and right circular polarizations, which is
generally very small in interstellar conditions 
(with the exception of masers).
Successful non-maser detections of 
the Zeeman effect in molecular clouds
have only been carried out with HI, OH, and CN lines 
because these species have the
largest Zeeman splitting factors ($\sim$2 -- 3.3 Hz/$\mu$G) among all
molecular lines (Crutcher et al.\ 1996, 1999; Falgarone et al.\ 2008).
Thermal HI and OH lines, however, probe relatively low-density gas
($n$(H) $< 10^4$ cm$^{-3}$).  Also, CN detections are difficult; 
Crutcher (2012) described only 8 CN Zeeman detections towards
14 positions observed with significant sensitivity.

ALMA Band 1 receivers provide the opportunity to detect the Zeeman
effect from the CCS 4$_{3}$--3$_{2}$ line at 45.37903 GHz and hence
greatly advance our understanding in star formation.  CCS has been
widely recognized as being present only very early in the star-forming
process through chemical models (Aikawa et al.\ 2001, 2005) and
observations (Suzuki et al.\ 1992; Lai \& Crutcher 2000). Therefore the
mass-to-flux ratio derived from the CCS Zeeman measurements will be
very close to the initial values before the onset of gravitational
collapse.  CCS 4$_{3}$--3$_{2}$ also has a relatively large Zeeman
splitting factor ($\sim$ 0.6 Hz/$\mu$G; Shinnaga \& Yamamoto 2000)
compared to most molecules.  ALMA's antennas and site will be excellent
at these ``long'' wavelengths, providing the stability and accuracy needed
for such sensitive polarization work. The linearly polarized detectors on
ALMA's antennas will also be ideally suited to measurement of Stokes
V signatures from CCS.

\begin{figure} 
\centering
\includegraphics[scale=0.8]{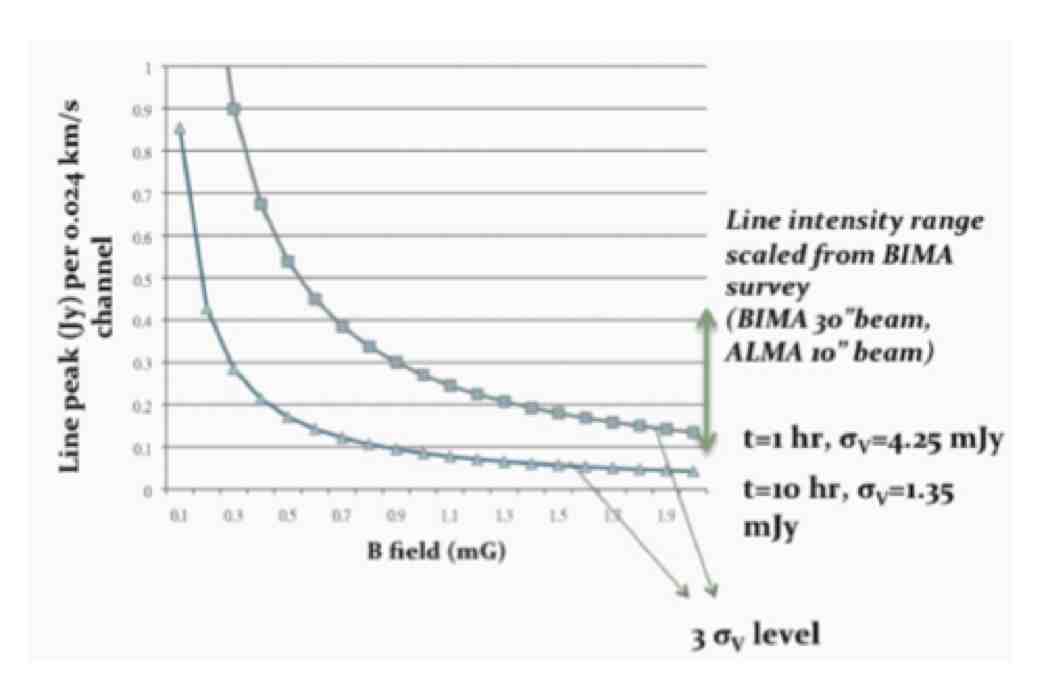}
\caption{The expected detection limits (3 $\sigma$) with integration time of 1 hr and 10 hr for a range of magnetic field strengths and CCS line intensity.}
\label{fig:ccs}
\end{figure}

Using the BIMA survey results from Lai \& Crutcher (2000), 
Figure~\ref{fig:ccs}
demonstrates that detections of CCS Zeeman effects can be achieved 
if the ALMA specifications for Band 1 receivers are met.
Zeeman effect
detection depends on two factors: the magnetic field strength and the line
intensity.  The two lines in Fig.~\ref{fig:ccs} show the 3 $\sigma$ detection
limits for Stokes V spectrum
with channel width of 0.024 km s$^{-1}$ and 1 hr or 10 hr integration time
for a range of magnetic field strengths and line intensities.  The channel
width is chosen to have at least 6 channels across the FWHM of the total
intensity spectrum (Stokes I).  If we scale the line intensity from Lai \&
Crutcher (2000) assuming the intensity distribution is uniform within
the 30$\arcsec$ BIMA beam, the expected line intensity would be around
0.1-0.4 Jy for ALMA observations with 10$\arcsec$ beam. Therefore, 
Fig.~\ref{fig:ccs} shows that for the magnetic fields of 0.2-1 mG (typical
values estimated from the application of the Chandrasehkar-Fermi method
to dust polarimetry in dense cores), we can detect the CCS Zeeman effect
with reasonable on-source integration time (less than 10 hr).

Note that the SiO v=1, J=1--0 transition at 43.12 GHz could be also used to
probe magnetic fields using the Zeeman effect, under certain circumstances.
Though its Zeeman splitting factor is lower than that of the CCS 4$_{3}$--3$_{2}$
line, the Zeeman effect may be detectible in situations where the SiO line is
extraordinarily bright, e.g., as a maser (see McIntosh, Predmore \& Patel
1994).  (Note, however, that non-Zeeman interpretations of circularly
polarized SiO emission have also been advanced; see Weibe \& Watson 1998).

In summary, ALMA Band 1 receivers will provide the opportunity to measure
the initial mass-to-flux ratio of molecular cores through the detection of the
Zeeman effect.  ALMA's linear feeds are ideally suited to measuring Stokes V and
ALMA's ability to recover extended, low surface brightness emission through the
shorter baselines of the 12-m Array and the inclusion of the ACA will be critical.
E.g., Roy et al.\ 2011 noted that the JVLA only recovered 1-13\% of the integrated
emission of CCS 2$_{1}$--1$_{0}$  observed in single-dish observations using
the JVLA's most compact (D) configuration.)  The results from Zeeman splitting
from ALMA will allow us to test realistically the expectations from theoretical
and numerical models for the first time.

\subsubsection{Molecular Outflows from Young Stars}

The Submillimeter Array (SMA) has proven to be a successful instrument for the study of the youngest molecular outflows 
and jets from the most deeply embedded sources (e.g., Hirano et al.\ 2006; Palau et al.\ 2006; Lee et al.\ 2007a,b, 2008, 
2009). The detection of excitation from rotational transitions of SiO up to levels $J$=8--7 and CO up to $J$=3--2 have 
uniquely identified a molecular high-velocity jet-like component located within outflow shells.  This component 
displays similarities to the optical forbidden line jets observed in T-Tauri stars (Hirano et al.\ 2006; Palau et al.\ 2006; 
Codella et al.\ 2007; Cabrit et al.\ 2007). These observations have provided a new probe of how jets are launched and 
collimated during the earliest protostellar phase. 

One unique opportunity offered by the Band 1 frequency range is observation of the $J$=1--0 transition of the SiO 
molecule at 43.424 GHz.  This transition has not yet been detected nor surveyed around even the brightest molecular 
outflows, except using single-dish telescopes (Haschick \& Ho 1990). One feature of this line that may be potentially
distinct from the higher-$J$\ transitions of SiO is that it may be tracing the outer and more diffuse gas located on the 
outskirts of outflow shells that can be easily excited by shocks. Potential morphological and kinematic studies of the
regions where the outflows interact with their own pre-natal clouds could be contrasted with other transitions using 
knowledge of their excitation conditions.  In particular, the improved sensitivity to extended emission and higher image
fidelity of ALMA make observations of SiO $J$=1--0 toward outflows more attractive with ALMA than with the JVLA.

\subsubsection{Co-Evolution of Star Formation and Active Galactic Nuclei}

Roughly half of the high-redshift objects detected in CO
line emission are believed to host an active galactic nucleus (AGN). 
Although they are selected based on their AGN properties, optically 
luminous high-redshift quasars exhibit many characteristics
indicative of ongoing star formation, 
e.g., thermal emission from warm dust (Wang et al.\ 2008) or extended
UV continuum emission. Indeed, galaxies with AGNs in the local Universe 
reveal a strong correlation between the mass ($m$) in their supermassive
black hole (SMBH) and that of 
their stellar bulge (measured from the stellar velocity dispersion ($\sigma$); 
e.g., Kormendy \& Richstone 1995; Magorrian et al.\ 1998; Gebhardt et
al.\ 2000).  Such a correlation can be explained if the SMBH formed
coevally with the stellar bulge, implying that the  
luminous quasar activity signaling the formation of a sub-arcsecond SMBH at
high-redshift should be accompanied
by starburst activity. High spatial resolution observations of CO
line emission in high-redshift quasars 
can be used to infer the dynamical masses, which are found to be comparable
to the derived molecular gas + black hole masses, meaning that their
stellar component cannot contribute 
a large fraction of the total mass. 

There is mounting evidence that 
quasar host galaxies at redshifts $z$ = 4--6 have SMBH masses
up to an order of magnitude larger 
than those expected from their bulge masses and the local relation 
(Walter et al.\ 2004; Riechers et al., in prep.),
suggesting that the SMBH may have formed first.
The possible time evolution of the $m - \sigma$ relation is of
fundamental importance in studies of galaxy evolution, and 
this new finding needs to be made more statistically robust.
Future observations of high-redshift AGN with the Band 1 receivers
on ALMA would allow us to address this question through the study of 
low-$J$\ CO line emission in galaxies beyond redshifts $z \approx1.3$
(see \S4.2).  ALMA especially allows studies of examples of such
objects in the south that are not well observable (if at all) with the JVLA.

\subsubsection{The Molecular Gas Content of Star-Forming Galaxies at $z \sim 2$}

While low-$J$\/ CO line emission has only been detected in a few high-redshift objects, 
high-$J$\/ CO line emission has been detected in more than sixty sources, most of which are classified as either 
submillimeter galaxies (SMGs) or far-infrared (FIR) luminous QSOs (see Carilli et al.\ 2011 for a review). Most of these 
studies have been conducted with sensitive interferometers and single-dish facilities operating in the 3~mm band (e.g., 
ALMA Band 3), which is sensitive to higher-$J$ CO line transitions at high redshift, as is illustrated in Figure~\ref{fig:red2}. 
These lines generally trace warmer and denser gas, and so previous data may have led to a bias in our understanding of 
the molecular gas properties of high-redshift galaxies (e.g., Papadopoulos \& Ivison 2002). The addition of Band 1 receivers 
on ALMA will allow comparisons of the cold gas traced by the low-$J$\/ transitions ($J$=2--1/1--0) in galaxies from 
moderate redshifts ($z \approx 1.3$) to those which existed when the Universe was re-ionized sometime before $z \ga 6$.

Although many previous studies of CO line emission in high-redshift
galaxies have focused on those starburst galaxies and AGN undergoing
episodes of extreme star formation (e.g., $\gg$100~M$_{\odot}$~yr$^{-1}$),
significant masses of molecular gas ($>10^{10}$~M$_{\odot}$) have been
discovered in more modest star-forming galaxies at $z = 1.5 - 2.0$
(Daddi et al.\ 2008). These ``BzK'' galaxies are selected for their
location in a B-$z$-K colour diagram (Daddi et al.\ 2004) and have
star-formation rates of $\sim$100~M$_{\odot}$~yr$^{-1}$ (Daddi et
al.\ 2007), while their number density is roughly a factor of 30
larger than that of the more extreme SMGs at similar
redshifts. Observations of CO $J$=2--1 line emission in these BzK galaxies
reveal comparable masses of molecular gas to that of the SMGs, so
their star-formation efficiencies appear lower. The excitation
conditions of their molecular gas (temperature and density) are
similar to those of the Milky Way (Dannerbauer et al.\ 2008), as
indicated by the ``turnover'' in the CO line spectral energy
distribution occurring at the {J}=3--2 transition, i.e., lower than
that of the SMGs which typically occurs at the {J}=6--5 or {J}=5--4
transition (Weiss et al.\ 2005).  To develop a full spectral energy
distribution for the CO line excitation, observations of these
galaxies in the $J$=1--0 transition are needed with Band 1 receivers on
ALMA.  Such data will also provide a more robust
estimate of the total molecular gas mass, along with the spatial
resolution needed to constrain the gas kinematics, as has been
done for the SMGs (Tacconi et al.\ 2006).  Indeed, recent high-resolution
studies of CO $J$=1--0 from lensed Lyman Break galaxies (Riechers
et al.\ 2010) and unlensed BzK galaxies (Aravena et al., in prep.)\/
have been made with the JVLA.  Also, CO $J$=1--0 emission has
been detected with the JVLA or GBT towards SMGs Ivison et al.\ 2010,
2011; Frayer et al.\ 2011; Riechers et al.\ 2011a,b).  ALMA observations
will allow similar important investigations to occur towards southern 
objects, especially those traced by ALMA itself in its higher-frequency
Bands.

\newpage
\section{Summary}
\label{sec:conc}

The Band 1 receiver suite has been considered an essential part of ALMA from the earliest planning days.  Even through the re-baselining exercise in 2001, the importance of Band 1 was emphasized. With the ALMA Development Plan underway, we have undertaken an updated review of the scientific opportunity at these longer wavelengths.  This document presents a set of compelling science cases over this frequency range.  The science cases reflect the new proposed range of Band 1, 35-50 GHz (nominal) with an extension up to 52 GHz, which was in fact chosen to optimize the science return from Band 1.  The science cases range from nearby stars and galaxies to the re-ionization edge of the Universe.  Two provide additional leverage on the present ALMA Level One Science Goals and are seen as particularly powerful motivations for building the Band 1 receiver suite: (1) detailing the evolution of grains in protoplanetary disks, as a complement to the gas kinematics, requires continuum observations out to $\sim 35\,$GHz ($\sim 9\,$mm); and (2) detecting CO 3 -- 2 spectral line emission from Galaxies like the Milky Way during the era of re-ionization, $6 < z < 10$
also requires Band 1 receiver coverage.  Band 1 receivers will also allow the pursuit of a diverse range of science 
cases that take advantage of the ALMA's particular strengths over other facilities (e.g., the JVLA).


\newpage

\section{References}
\label{sec:refs}

\noindent
Acke, B., \& van den Ancker, M. E.\ 2004, A\&A, 426, 151

\noindent
Acke, B., et al.\ 2004, A\&A, 422, 621

\noindent
Adams, F. C., Emerson, J. P., \& Fuller, G. A.\ 1990, ApJ, 357, 606


\noindent
Anglada, G.\ 1995, RMAACS, 1, 67

\noindent
Aikawa, Y., Ohashi, N., Inutsuka, S.-I., Herbst, E., \& Takakuwa, S.\ 2001, \apj, 552, 639

\noindent
Aikawa, Y., Herbst, E., Roberts, H., \& Caselli, P.\ 2005, \apj, 620, 330

%

\noindent
Andrews, S., \& Williams, J. P.\ 2007a, ApJ, 659, 705

\noindent
Andrews, S., \& Williams, J. P.\ 2007b, ApJ, 671, 1800


\noindent
Aschwanden, M. J., Wills, M. J.; Hudson, H. S., Kosugi, T., Schwartz, R. A.\ 1996, \apj, 468, 398

\noindent
Bailey, J., Chrysostomou, A., Hough, J. H., Gledhill, T. M., McCall, A., Clark, S., Menard, F., \& Tamura, M. 1998, Science, 281, 672

\noindent
Bartel, N.\ et al.\ 2002, ApJ, 581, 404

\noindent
Bartel, N., Bietenholz, M.F., Rupen, M.P., \& Dwarkadas, V.V.\ 2007, ApJ, 668, 924

\noindent
Bauer, F. E., et al.\ 2008, ApJ, 688, 1210

\noindent
Beckwith, S., Sargent, A. I., Chini, R. S., \& Guesten, R.\ 1990, AJ, 99, 924

\noindent
Beckwith, S., \& Sargent, A. I.\ 1991, ApJ, 381, 250

\noindent
Bertoldi, F.\ et al.\ 2003, A\&A, 409, L47

\noindent
Bietenholz, M. F., Bartel, N., \& Rupen, M. P.\ 2004, Science, 304, 1947

\noindent
Bietenholz, M. F., Bartel, N., \& Rupen, M. P.\ 2010, ApJ, 712, 1057

\noindent
Blain, A.~W., Smail, I., Ivison, R.~J., Kneib, J.-P., \& Frayer, D.~T.\ 2002, \physrep, 369, 111 

\noindent
Boley, A., et al. 2012, \apj, 750, L21

\noindent
Bonamente, M., Joy, M., LaRoque, S.~J., et al.\ 2008, \apj, 675, 106 

\noindent
Birkinshaw, M.\ 1999, Phys. Rep.\ 310, 97

\noindent
Buckle, J. V., et al.\ 2006, Faraday Discussions, 133, 63

\noindent
Bouwens, R. J., et al.\  2009, \apj, 690, 1764

\noindent
Boss, A.\  2005, ApJ, 629, 535

\noindent
Butler, B., 2010, VLA Test Memo 232, ``Atmospheric Opacity at the VLA"

\noindent
Butler, B., 1998, VLA Memo 237, ``Precipitable Water at the VLA -- 1990-1998"

\noindent
Butler, B. \& Desai, K., VLA Test Memo 222, ``Phase Fluctuations at the VLA Derived From
One Year of Site Testing Interferometer Data"

\noindent
Cabrit, S., et al.\ 2007, \aap, 468, L29

\noindent
Calvet, N., et al.\  2002, ApJ, 568, 1008

\noindent
Cappelluti, N., Predehl, P., B{\"o}hringer, H., et al.\ 2011, Memorie della Societa Astronomica Italiana 
Supplementi, 17, 159 

\noindent
Carilli, C. L., et al.\ 2007, ApJ, 666, L9

\noindent
Carilli, C. L., et al.\ 2008, Ap\&SS, 313, 307

\noindent
Carlstrom, J. E., Holder, G. P., \& Reese, E. D.\ 2002, \araa, 40, 643

\noindent
Casassus, S., et al.\ 2008, MNRAS, 391, 1075

\noindent
Chevalier, R.A.\ 1982, ApJ, 259, 85

\noindent
Churchwell, E. 2001, ARA\&A, 40, 27

\noindent
Cohen, R. J.\ 1995, Ap\&SS, 224, 55

\noindent
Codella, C., et al.\ 2007, \aap, 462, L53

\noindent
Cool, R.J., et al.\  2006, \aj, 132, 823


\noindent
Cragg, D. M., Sobolev, A. M., \& Godfrey, P. D.\ 2005, MNRAS, 360, 533

\noindent
Crutcher, R. M. 2012, ARA\&A, 50, 29

\noindent
Crutcher, R. M., Troland, T. H., Lazareff, B., \& Kazes, I.\ 1996, \apj, 456, 217

\noindent
Crutcher, R. M., Troland, T. H., Lazareff, B., Paubert, G., \& Kaz{\`e}s, I.\ 1999, ApJL, 514, 121

\noindent
D'Addario, L., \& Holdaway, M. 2003, ALMA Memo 521, ``Joint Distribution of Atmospheric 
Transparency and Phase Fluctuations at Chatnantor"

\noindent
Daddi, E., et al.\ 2004, ApJ, 617, 746 

\noindent
Daddi, E., et al.\ 2007, ApJ, 670, 156 

\noindent
Daddi, E., et al.\ 2008, ApJL, 673, L21 

\noindent
Dannerbauer, H., et al.\ 2009, ApJL, 698, 178 

\noindent
Dark Energy Survey Collaboration, The 2005, arXiv:astro-ph/0510346 

\noindent
Dib, S., Kim, J., V{\'a}zquez-Semadeni, E., Burkert, A., \& Shadmehri, M.\ 2007, \apj, 661, 262

\noindent
Dodds-Eden, K., et al.\ 2009, \apj, 698, 676

\noindent
Draine, B.\  2003, ARA\&A, 41, 241

\noindent
Draine, B.\  2006, ApJ, 636, 1114

\noindent
Draine, B., \& Anderson, N.\ 1985, ApJ, 292, 494 

\noindent
Draine, B., \& Lazarian, A.\ 1998, ApJ, 508, 157

\noindent
Dullemond, C. P., \& Dominik, C.\ 2005, A\&A, 434, 971

\noindent
Dunkley, J., et al.\ 2009, ApJS, 180, 306

\noindent
Dunkley, J., Hlozek,  R., Sievers, J., et al.\ 2011, \apj, 739, 52 

\noindent
Eckart, A., et al.\ 2006, \aa, 450, 535

\noindent
Ellingsen, S. P.\ et al.\  2007, IAUS, 242, 213

\noindent
Ellingsen, S. P.\  2004, IAUS, 221, 133

\noindent
Ettori, S., \& Fabian, A.~C.\ 2006, \mnras, 369, L42 

\noindent
Falgarone, E., Troland, T. H., Crutcher, R. M., \& Paubert, G.\ 2008, \aap, 487, 247

\noindent
Fan, X., et al.\ 2001, AJ, 122, 2833

\noindent
Fan, X., et al.\  2006a, \aj, 132, 117

\noindent
Fan, X., Carilli, C. L., \& Keating, B.\  2006b, \araa, 44, 415

\noindent
Fender, R. P., et al.\ 1999, ApJL, 519, 165

\noindent
Finkbeiner, D. P., Schlegel, D. J., Frank, C., \& Heiles, C.\ 2002, ApJ, 566, 898

\noindent
Frayer, D., et al.\ 2011, \apj, 726, L22

\noindent
Gaensler, B. M., et al.\ 2007, AIP Confer. Ser., ed.\ S.\ Immler \& R. McCray, 937, 86

\noindent
Garrod, R. T., Weaver, S. L. W., \& Herbst, E.\ 2008, \apj, 682, 283

\noindent
Gebhardt, K., et al.\ 2000, ApJ, 543, L5

\noindent
Geisbuesch, J., \& Hobson, M. P.\ 2007, \mnras, 382, 158

\noindent
Geisbuesch, J., Kneissl, R., \& Hobson, M. P.\ 2005, \mnras, 360, 41

\noindent
Ghez, A., et al. 2008, \apj, 689, 1044

\noindent
Gillessen, S., et al.\ 2009, \apj, 692, 1075

\noindent
Glikman, E., et al.\  2008, \aj, 136, 954

\noindent
Goldreich, P., Keeley, D. A., \& Kwan, J. Y.\ 1973, ApJ, 179, 111


\noindent
Greaves, J., Richards, A. M. S., Rice, W. K. M., \& Muxlow, T. W. B.\ 2008, MNRAS, 391, 74

\noindent
Guilloteau, S., et al. 20009, \aa, 529, 105

\noindent
Gunn, J.E., \& Peterson, B.A.\  1965, \apj, 142, 16331


\noindent
Hales, A. S., et al.\   2004, ApJ, 613, 977


\noindent
Haschick, A. D., \& Ho, P. T. P.\ 1990, \apj, 352

\noindent
Harvey-Smith, L., Soria-Ruiz, R., Duarte-Cabral, A., \& Cohen, R. J.\ 2008, MRAS, 384, 719 

\noindent
Helfand, D. J., Gotthelf, E. V., \& Halpern, J. P.\ 2001, ApJ, 556, 380

\noindent
Hincks, A. D., et al.\ 2009, arXiv:0907.0461

\noindent
Hirano, N., et al.\ 2006, ApJL, 636, 141

\noindent
Holtom, P. D., Bennett, C. J., Osamura, Y., Mason, N. J, \& Kaiser, R. I.\ 2005, \apj, 626, 940

\noindent
Hu, E.,M., Cowie, L. L., \& McMahon, R. G.\ 1998, ApJ, 502, L99

\noindent
Hunt-Cunningham, M. R., \& Jones, P. A.\ 2004, IAUS, 213, 159

\noindent
Isella, A, Natta, A., Wilner, D., Carpenter, J. M.. \& Testi, L.\ 2010, \apj, 725, 1735

\noindent
Ivison, R. J., et al.\ 2010, MNRAS, 404, 198

\noindent
Ivison, R. J., et al.\ 2011, MNRAS, 412, 1913

\noindent
Iye, M., et al.\ 2006, Nature, 443, 186

\noindent
Kaufmann, P., Raulin, J.-P., de Castro, C. G. G.\ 2004, \apj, 603, L121

\noindent
Kepley, A. A., Chomiuk, L., Johnson, K. E., Goss, W. M., Balser, D. S., \& Pisano, D. J. 2011, ApJ, 739, L24

\noindent
Kitayama,T., Komatsu, E., Ota, N., Kuwabara, T., Suto, Y., Yoshikawa, K., Hattori, M., \& Matsuo, H. 2004, \pasj, 56, 17

\noindent
Koerding E. G., Fender R. P., \& Migliari S.\ 2006, MNRAS, 369, 1451

\noindent
Komatsu, E., Matsuo, H., Kitayama, T., Kawabe, R., Kuno, N., Schindler, S., \& Yoshikawa, K. 2001, \pasj, 53, 57

\noindent
Kormendy, J., \& Richstone, D.\ 1995, \araa, 33, 581

\noindent
Korngut, P.~M., Dicker,  S.~R., Reese, E.~D., et al.\ 2011, \apj, 734, 10 

\noindent
Lai, S.-P., \& Crutcher, R.M.\ 2000, \apjs, 128, 271

\noindent
Laki{\'c}evi{\'c}, M., Zarnado, G., van Loon, Th., Staveley-Smith, L., Potter, T., Ng, C.-Y., \& Gaensler, B.~M. 2012, A\&A, 541, L2 

\noindent
Lee, C.-F., et al.\  2007a, \apj, 659, 499

\noindent
Lee, C.-F., et al.\ 2007b, \apj, 670, 1188

\noindent
Lee, C.-F., et al.\ 2008, \apj, 685, 1026

\noindent
Lee, C.-F., et al.\ 2009, \apj, 699, 1584

\noindent
Lee, J., \& Komatsu, E.\ 2010, \apj, 718, 60 

\noindent
Leger, A., \& Puget, J. L.\ 1984, \aap, 137, L5

\noindent
Li, Y., et al.\ 2007, ApJ, 665, 187

\noindent
Li, Y., et al.\ 2008, ApJ, 678, 41

\noindent
Lim, J., \& Takakuwa, S.\ 2006, \apj, 653, 425

\noindent
Lonsdale, C. J., et al.\ 2006, ApJ, 647, 185

\noindent
Maccarone T. J., 2008, ASPC, 401, 191

\noindent
Maccarone T. J., Gallo E., Fender R. P.\ 2003, MNRAS, 345, L19

\noindent
Magorrian, J., et al.\ 1998, AJ, 115, 2285

\noindent
Maness, H.\ et al.\ 2008, ApJ, 686, L25

\noindent
Mannings, V., \& Emerson, J.\ 1994, MNRAS, 267, 361

\noindent
Markevtich, M., et al.\ 2000, \apj, 541, 542

\noindent
Markevitch, M., et al.\ 2002, \apj, 567, L27


\noindent
Markevitch, M., et al.\ 2009, arXiv:0902.3709

\noindent
Mason, B.~S., Dicker,  S.~R., Korngut, P.~M., et al.\ 2010, \apj, 716, 739 

\noindent
Masuda, S., Kosugi, T., Hara, H., Tsuneta, S., Ogawara, Y.\ 1994, Nature, 371, 495


\noindent
McIntosh, G. C., Read Predmore, C., \& Patel, N. A. 1994, \apj, 428, L29

\noindent
McKee, C. F. \& Ostriker, E. C. 2007, ARA\&A, 45, 565

\noindent
Melis, C., et al. 2011, \apj, 739, L7

\noindent
Menanteau, F., Hughes, J.~P., Sifon, C., et al.\ 2011, arXiv:1109.0953 

\noindent
Menten, K.\ 2007, IAUS, 242

\noindent
Merloni, A., et al. 2012, arXiv, 1209.3114

\noindent
Migliari S., Fender R. P., Rupen, M., Wachter, S., Jonker, P. G., Homan, J., \& van der Klis, M.\ 2004, MNRAS, 351, 186

\noindent
Morandi, A., Ettori, S., \& Moscardini, L.\ 2007, \mnras, 379, 518 


\noindent
Mortlock, D. J., et al.\ 2008, arXiv0810.4180

\noindent
Mouschovias, T.Ch.\  \& Spitzer, L., Jr.\ 1976, ApJ, 210, 326


\noindent
Narayanan, D., et al.\ 2008, ApJS, 174, 13

\noindent
Narayan, R., Ozel, F., \& Sironi, L. 2012, \apj, 757, L20

\noindent
Narayan R., \& Yi, I.\ 1994, ApJL, 428, 13

\noindent
Neufeld, D., \& Melnick, G.\ 1991, ApJ, 368, 215

\noindent
Ota, K., et al.\  2008, \apj, 677, 12

\noindent
Ot\'arola, A., Holdaway, M., Nyman, L-\AA, Radford, S. J. E., \& Butler, B. J. 2005, ALMA Memo 512,
``Atmospheric Transparency at Chajnantor: 1973-2003"

\noindent
Palau, A., et al.\ 2006,  ApJL, 636, 137

\noindent
Papadopoulos, P. P. \& Ivison, R. J.\ 2002, ApJ, 564, L9 

\noindent
Pascucci, I., Gorti, U., \& Hollenbach, D. 2012, ApJ, 751, L42

\noindent
Peters, T., Longmore, S. N., \& Dullemond, C. P. 2012, MNRAS, 425, 2352

\noindent
Pfrommer, C., En{\ss}lin, T.~A., \& Sarazin, C.~L.\ 2005, \aap, 430, 799 

\noindent
Plagge, T., et al.\ 2012, arXiv:1203.2175

\noindent
Planck Collaboration, Ade, P.~A.~R., Aghanim, N., et al.\ 2011, arXiv:1101.2024 

\noindent
Pinte, C., Menard, F., Duch\^{e}ne, G., \& Bastien, P.\ 2006, A\&A, 459, 797

\noindent
Pinte, C., et al.\  2009, A\&A, 498, 967

\noindent
Pollack, J., et al.\ 1996, Icarus, 124, 62

\noindent
Poole, G.~B., Babul, A., McCarthy, I.~G., et al.\ 2007, \mnras, 380, 437 


\noindent
Rafikov, R.\ 2006, ApJ, 646, 288

\noindent
Reid, M. J. \& Brunthaler, A. 2004, \apj, 616, 872

\noindent
Reipurth, B., \& Bally, J.\ 2001, \araa, 39, 403

\noindent
Rephaeli, Y.\ 1995, \araa, 33, 541

\noindent
Rhoads J. E., et al.\ 2000,  ApJ, 545, L85

\noindent
Ricci, L., Testi, L., Natta, A., Neri, R., Cabrit, S., \& Herczeg, G. 2010, \aa, 512, 15

\noindent
Riechers, D., et al.\ 2009, \apj, 703, 1338

\noindent
Riechers, D., et al.\ 2010, \apj, 724, L153

\noindent
Riechers, D., et al.\ 2011a, \apj, 733, L11

\noindent
Riechers, D., et al.\ 2011b, \apj, 739, L31

\noindent
Rodmann, J., et al.\ 2006, A\&A, 446, 211

\noindent
Rodr{\'{\i}}guez, L. F., et al.\ 1998, \nat, 395, 355

\noindent
Rodr{\'{\i}}guez, L. F., et al.\ 2003, ApJL, 586, 137

\noindent
Roy, N., Datta, A., Momjian, E. \& Sarma, A. P. 2011, ApJ, 739, L4

\noindent
Sarazin, C.L.\ 1988, in ``X-ray emission  from clusters of galaxies," Cambridge University Press

\noindent
Scaife, A. M. M., et al.\ 2009, MNRAS, 400, 1394

\noindent
Scaife, A. M. M., et al.\ 2010, MNRAS, 403, 46

\noindent
Sehgal, N., Trac, H., Acquaviva, V., et al.\ 2011, \apj, 732, 44 

\noindent
Shang, H., Lizano, S., Glassgold, A., \& Shu, F.\ 2004, ApJL, 612, 69

\noindent
Sheperd, D. S., Chandler, C. J., Cotton, B., Perley, R. A., Randall, S. K., \& Remijan, A. J. 2013, High Energy Density Physics, 9, 26

\noindent
Shinnaga, H., \& Yamamoto, S.\ 2000, \apj, 544, 330

\noindent
Shu, F. H., Adams, F. C., \& Lizano, S.\ 1987, \araa, 25, 23

\noindent
Sobolev, A. M., et al.\ 1999, in The Physics and Chemistry of the Interstellar Medium, ed. V. 
Ossenkopf et al.\ (GCA-Verlag: Herdecke), 299 

\noindent
Sobolev, A. M., Cragg, D. M., Godfrey, P. D.\ 1997, MNRAS, 288, L39


\noindent
Staniszewski, Z., et al.\ 2009, \apj, 701. 32

\noindent
Sui, L., Holman, G. D.\ 2003, \apj, 596, L251

\noindent
Sunyaev, R. A., \& Zel'dovich, Ya. B.\ 1972, Comm. Astrophys. Space  Phys. 4, 173

\noindent
Sutton, E. C., et al.\  2001, ApJ, 554, 173

\noindent
Suzuki, H.,  et al.\ 1992, ApJ, 392, 551

\noindent
Tacconi L. J., et al.\ 2006, ApJ, 640, 228

\noindent
Takakuwa, S., Kamazaki, T., Saito, M., \& Hirano, N.\ 2003, ApJ, 584, 818

\noindent
Tananbaum H., Gursky H., Kellogg E., Giacconi R., \& Jones C.\ 1972, ApJ, 177, L5

\noindent
Taniguchi, Y., et al.\ 2005, PASJ, 57, 165

\noindent
Testi, L., Natta, A., Shepherd, D.S., \& Wilner, D.J.\ 2001, ApJ, 554, 1087

\noindent
Tudose, V., Fender, R. P., Linares, M., Maitra, D., \& van der Klis M.\ 2009, MNRAS, 400, 2111

\noindent
Vikhlinin, A., Kravtsov, A.~V., Burenin, R.~A., et al.\ 2009, \apj, 692, 1060 

\noindent
Vlemmings, W. H. T., Diamond, P. J., \& Imai, H.\  2006, IAUS, 234, 267

\noindent
Wagg, J., \& Kanekar, N.\ 2012, \apj, 751, L24

\noindent
Wagg, J., Kanekar, N., \& Carilli, C.L.\ 2009, ApJL, 697, L33

\noindent
Walter, F., et al.\ 2003, Nature, 424, 406

\noindent
Walter, F., et al.\  2004, \apjl, 615, L17

\noindent
Walter, F., et al.\  2009, \apjl, 691, L1

\noindent
Wang, R., et al.\  2008, AJ, 135, 1201

\noindent
Wang, R., et al.\  2010, \apj, 714, 699

\noindent
Wang, R., et al.\ 2011a, \aj, 142, 101

\noindent
Wang, R., et al.\ 2011b, \apj, 739, L34

\noindent
Watson, W. D.\ 2008, arXiv0811,1292W

\noindent
Weintraub, D. A., Sandell, G., \& Duncan, W. D.\ 1989, ApJ, 340, 69

\noindent
Weiss, A., Downes, D., Walter, F., \& Henkel, C.\ 2005, A\&A, 440, 45

\noindent
Welch, W. J., et al.\ 2004, in Bioastronomy 2002, eds R. Norris, F. Stootman (ASP, San Francisco), 213, 59

\noindent
Weiler, K. W., Panagia, N., Montes, M. J., \& Sramek, R. A. 2002, ARA\&A, 40, 387


\noindent
Weibe, D. S., \& Watson, W. D. 1998, ApJ, 503, L71

\noindent
White, S. M., Krucker, S., Shibasaki, K., et al.\ 2003, \apj, 595, L111

\noindent
Wik, D.~R., Sarazin, C.~L.,  Ricker, P.~M., \& Randall, S.~W.\ 2008, \apj, 680, 17 

\noindent
Williamson, R., Benson, B.~A., High, F.~W., et al.\ 2011, \apj, 738, 139 

\noindent
Willott, C. J., et al.\  2009, \aj, 137, 3541

\noindent
Wilner, D. J., Ho, P. T. P., Kastner, J. H., \& Rodriguez, L. F.\ 2000, ApJ, 534, 101

\noindent
Wootten, A.\  2007, IAUS, 242

\noindent
Wyatt, M. C., 2009, ARA\&A, 46, 339

\noindent
Yamada, K., et al.\ 2012, PASJ, in press.

\noindent
Yokoyama, T., Akita, K., Morimoto, T., Inoue, K., Newmark, J.\ 2001, \apj, 546, L69

\noindent
Yusef-Zadeh, F., et al. 2012, \apj, 757, L1

\noindent
Yusef-Zadeh, F., et al. 2006, \apj, 650, 189

\end{document}